\documentclass[%
reprint,
superscriptaddress,
amsmath,amssymb,
aps,
pra,
]{revtex4-1}

\usepackage{graphicx}
\usepackage{dcolumn}
\usepackage{bm}
\usepackage[dvipsnames]{xcolor}

\usepackage{tabularx}
\newcolumntype{Y}{>{\centering\arraybackslash}X}
\newcommand{\tOmega}{\tilde{\Omega}}

\newcommand{\efield}{{\bm{\mathcal{E}}}}

\begin{document}

\title{Resonant and first-order dipolar interactions between ultracold molecules in static and microwave electric fields}
\author{Tijs Karman}
\affiliation{Radboud University, Institute for Molecules and Materials, Heijendaalseweg 135, 6525 AJ Nijmegen, the Netherlands}
\author{Zoe Z. Yan}
\affiliation{Department of Physics, Princeton University, Princeton, New Jersey 08544, United States}
\author{Martin Zwierlein}
\affiliation{MIT-Harvard Center for Ultracold Atoms, Research Laboratory of Electronics, and Department of Physics, Massachusetts Institute of Technology, Cambridge, Massachusetts 02139, USA}
\date{\today}
\begin{abstract}
We theoretically study collisions between ultracold polar molecules that are polarized by microwave or static electric fields.
We systematically study the dependence on field strength, microwave polarization, and detuning from rotational transitions.
We calculate the loss in two-body collisions that is observable experimentally and compare to the results expected for purely first-order dipolar interactions.
For ground state molecules polarized by a static electric field,
the dynamics are accurately described by first-order dipolar interactions.
For microwave dressing, instead, resonant dipolar collisions dominate the collision process, in which molecules reorient along the intermolecular axis and interact with the full strength of the transition dipole. For red detuning, reorientation can only be suppressed at extreme Rabi frequencies.
For blue detuned microwaves,
resonant dipolar interactions dominate even for high Rabi frequencies,
leading to microwave shielding for circular polarization and structured losses due to resonances for linear polarization.
The results are presented numerically for fermionic $^{23}$Na$^{40}$K and bosonic $^{23}$Na$^{39}$K molecules.
\end{abstract}

\maketitle

\section{Introduction}

Ultracold polar molecules provide a promising platform for quantum science and technology.
They can be produced either by direct laser cooling of molecules~\cite{truppe:17,anderegg:18},
or by associating pairs of ultracold atoms~\cite{ni:08, lang:08,danzl:10,takekoshi:14, molony:14, park:15, guo:16,rvachov:17, seesselberg:18}.
Many proposed applications of ultracold molecules rely on their tunable long-range dipole-dipole interactions~\cite{buchler:07, micheli:06,micheli:07,bruun:08, cooper:09,wall:10, gorshkov:11, levinsen:11}.
In the absence of applied fields the molecules' dipole moment is not oriented.
Inducing non-zero dipole moment in the lab frame requires external fields that mix even and odd rotational states.
This can be accomplished using either static electric fields~\cite{ni:10,guo:18},
or microwave fields tuned close to a rotational transition~\cite{yan:20}.

In Ref.~\cite{yan:20}, we have presented a combined experimental and theoretical study of resonant dipolar collisions induced by dressing ultracold fermionic $^{23}$Na$^{40}$K molecules with microwaves red-detuned from the lowest rotational transition, a 5.6~GHz splitting, with Rabi frequencies on the order of kHz and no active control of polarization.
In this paper, we perform a systematic theoretical study of the dependence on field strength, detuning, and polarization.
We explore the two-body collisional dynamics as a function of induced dipole moment,
and compare observable loss rates for molecules polarized using static or microwave electric fields.
We use coupled-channels (c.c.) scattering calculations with a fully absorbing boundary condition at short range.
This boundary condition has been shown to accurately reproduce experimental loss rates even for \emph{nonreactive} ultracold molecules~\cite{ye:18, gregory:19}.
The physical origin of this short-range loss for nonreactive molecules was previously attributed to ``sticky collisions''~\cite{mayle:13},
but is now thought to be caused by excitation of collision complexes by the trapping laser~\cite{christianen:19b,christianen:19a,gregory:20,liu:20}.

\begin{figure*}
\begin{center}
\includegraphics[width=0.32\textwidth]{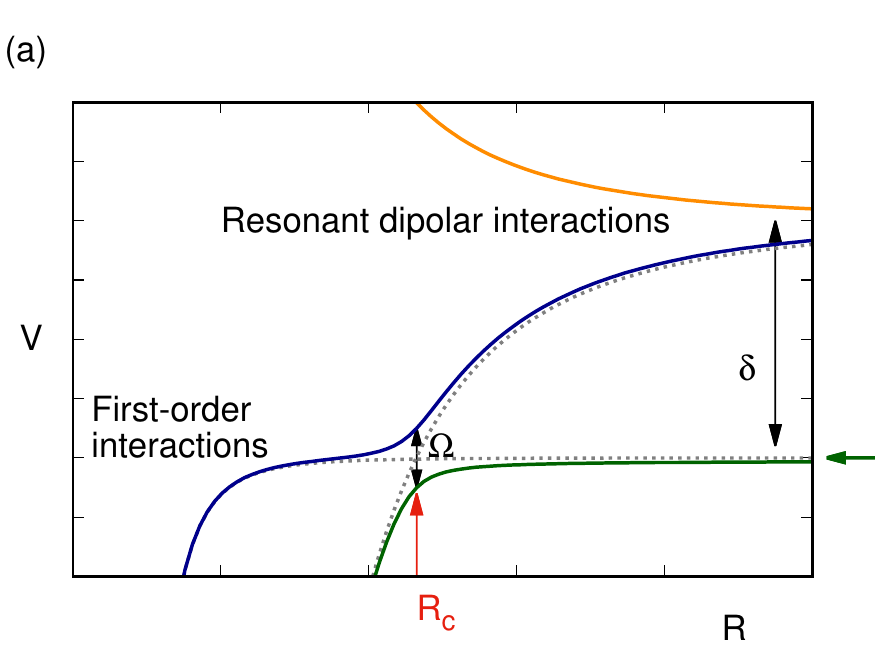}
\includegraphics[width=0.32\textwidth]{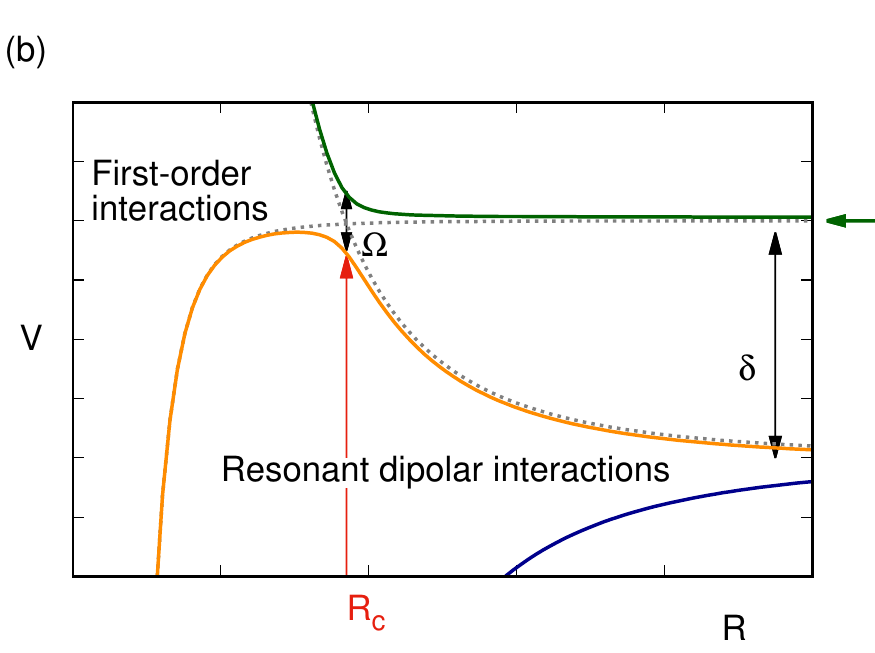}
\includegraphics[width=0.32\textwidth]{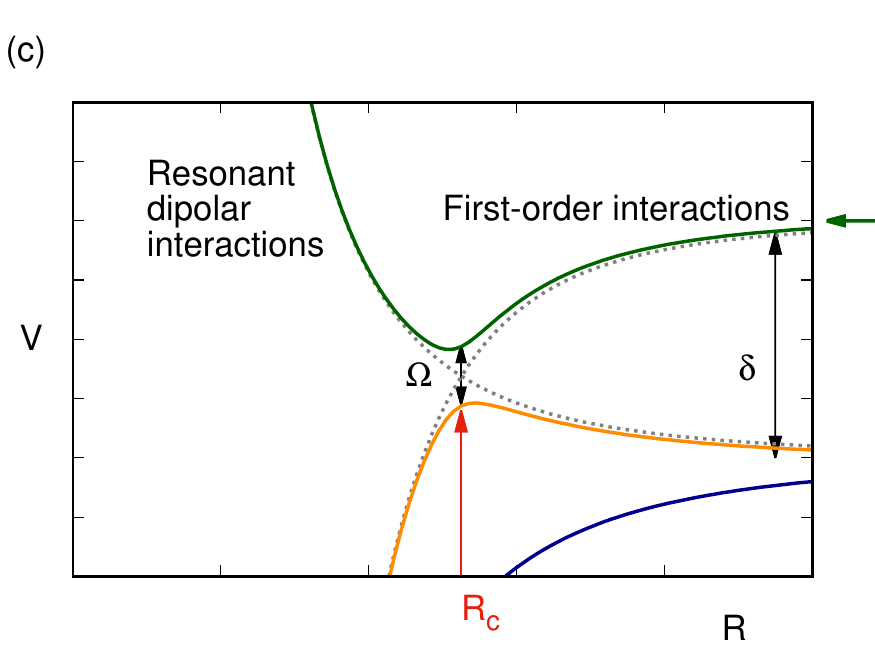}
\caption{ \label{fig:cartoon}
Illustration of the competition of first-order and resonant dipolar interactions explored here.
Dashed lines indicate diabatic interaction potentials,
which cross at the Condon point marked $R_c$,
whereas the solid lines indicate adiabatic potentials for which the crossing is avoided due to the Rabi coupling.
	Panel (a) sketches the interaction potentials for a pair of resonantly interacting states occurring above the initial state (in green, indicated by an arrow), as is realizable by dressing with microwaves red detuned from a rotational transition.
Here, the attractive branch of the resonant dipolar interaction crosses the ground state interaction potential,
where coupling between these potentials can increase collision rates by orders of magnitude~\cite{yan:20}.
In the illustration, the Condon point occurs outside the range of first-order dipolar interactions.
If by contrast the Condon point occurred at shorter range, the molecules would continue to short range regardless of whether they traverse this avoided crossing diabatically or adiabatically,
such that it is only the first-order interactions that determine the observable loss rates.
This is the scenario realized for ground-state molecules polarized by a static electric field.
Microwave dressing with blue detuning also gives access to resonantly interacting states below the initial state, shown in panel (b).
Here, the repulsive branch of the resonant dipolar interaction crosses the initial potential,
which can reduce short-range loss rates by orders of magnitude even if the interactions become repulsive only at shorter range, as shown in panel (c).
The combination of long-range first-order dipolar interactions and repulsive resonant dipolar interactions at shorter range can support quasi-bound states and lead to scattering resonances.
}
\end{center}
\end{figure*}

The qualitative physical picture we explore here is illustrated in Fig.~\ref{fig:cartoon}.
By applying an external field, either microwave or static electric, we induce a dipole moment in the ground-state molecules, leading to tunable first-order dipole-dipole interactions.
However, there also exist other states of the pair of molecules,
where each molecule is in a state of opposite parity such that a large transition dipole moment between these states leads to so-called resonant dipolar interactions.
The resonant dipolar interaction quantizes the molecular dipoles about the intermolecular axis,
a qualitatively different scenario from first-order dipolar interactions between dipole moments oriented along the external field.
The resulting interaction potential can cross the interaction potential for two ground-state molecules,
and coupling between these states at this Condon point can qualitatively change the dynamics, boosting or suppressing collision rates by orders of magnitude~\cite{karman:18d,yan:20}.

Whether such an avoided crossing controls the collision dynamics depends on the strength of the interactions at the Condon point compared to the collision energy or equivalently temperature,
which can be understood as follows.
The collision dynamics are determined by interactions where these become comparable to the kinetic energy,
which for ultracold collisions and long-ranged interactions can occur at very large intermolecular distances.
At larger distances the interactions can be neglected,
whereas at much shorter distances -- where the interactions are very strong -- any flux that reaches these distances will classically continue to short range.
Thus, for attractive resonant dipolar interactions, the avoided crossing controls the collision dynamics if the Condon point occurs at such large distance that the first-order interaction is weak compared to the collision energy or temperature,
see Fig.~\ref{fig:cartoon}(a).
If the Condon point occurs at much shorter distances,
it is the first-order interaction at larger distances that determines whether the molecules reach short range, regardless of whether they do this by subsequently crossing the Condon point diabatically or adiabatically.
For repulsive interactions, on the other hand, these interactions can prevent molecules from reaching short range, even if the Condon point occurs at much shorter distances,
see Fig.~\ref{fig:cartoon}(b,c).

For microwave dressing detuned by $\delta=\omega-2B_\mathrm{rot}/\hbar$ from a rotational transition,
a pair of resonantly interacting states occurs at energy $\hbar\delta$ \emph{below} the initial state, which corresponds to two ground-state molecules.
Using red detuning $\delta <0$ and blue detuning $\delta>0$, microwave dressing gives access to both attractive and repulsive resonant dipolar interactions.
The coupling at the Condon point is controlled by the Rabi frequency, $\Omega$.
The detuning, $\delta$, can be chosen small such that the Condon point occurs at long range, where it controls the dynamics.
For example, Ref.~\cite{yan:20} used kHz Rabi frequencies and detuning up to 40~kHz,
which leads to a Condon point at a separation of thousands of Bohr radii,
much larger than the range of van der Waals interactions that dominate in the absence of microwave dressing and act on a range of hundreds of Bohr radii. 

For ground-state molecules in a static electric field, 
pairs of resonantly interacting excited states necessarily occur above the initial state,
meaning that they cross the ground-state interaction potential only with attractive resonant interactions.
Furthermore, excited states lie higher in energy on a scale set by the rotational constant, which is on the order of GHz,
\emph{i.e.}\  tens of milikelvin,
which is much higher than the kHz to MHz regime accessible by microwave dressing.
This means that any avoided crossings occur at very short distances and for attractive resonant interactions,
and therefore do not control the dynamics.
The collision dynamics for ground-state molecules polarized by a static electric field are completely determined by the first-order dipolar interaction.

In order to access resonant interactions using static electric fields, molecules must be prepared excited states.
An example of this is the resonant shielding~\cite{quemener:16,gonzalez:17},
demonstrated recently~\cite{matsuda:20,li:21},
where molecules are prepared in the first rotational excited state,
and the electric field strength is varied to tune this initial state into resonance with the energy of the combination of one ground-state molecule $J=0$ and one molecule further excited to $J=2$.
We do not consider this explicitly in the present work.

After this initial qualitative description we discuss the dipole moment and dipole-dipole interactions induced by polarizing the molecules using static or microwave electric fields in Sec.~\ref{sec:dip}.
Next, the details of coupled-channels calculations of collision rates in Sec.~\ref{sec:theory}.
The resulting experimentally observable loss rates are presented and analyzed in Sec.~\ref{sec:loss},
and concluding remarks are given in Sec.~\ref{sec:conclusions}.

\section{Inducing dipole-dipole interactions \label{sec:dip}}

\subsection{Static electric fields}

In field-free space, the eigenstates of a molecule are the rotational eigenstates,
which have well-defined parity and hence zero expectation value of the dipole moment.
Hence, even polar molecules, \emph{i.e.}, molecules with an electric dipole moment, $d$, in their body-fixed frame of reference, have zero dipole moment in the laboratory frame.
Inducing dipole-dipole interactions therefore requires breaking inversion symmetry by applying an external field in which the molecule polarizes.
A static electric field achieves this through the Stark interaction $-\hat{\bm{d}} \cdot \efield$,
which mixes $J'=J\pm1$ rotational states.
This leads to an induced space-fixed dipole moment, $d_\mathrm{ind}$, that is tunable between zero and the molecule's body-fixed electric dipole moment, $d$, by increasing the electric field strength.
Polarization is achieved when $d E \sim B_\mathrm{rot}$, so for a field strength $E \sim B_\mathrm{rot}/d$.

A simple but relevant model of dipole-dipole interactions is the ``space-fixed dipole'' approximation~\cite{bohn:09},
\begin{align}
\hat{H}^{(\mathrm{SF~dipole})} = -\frac{\hbar^2}{2\mu} \nabla^2 - \frac{2 d_\mathrm{ind}^2}{4\pi\epsilon_0 R^3} P_2(\cos\theta),
\label{eq:BCT}
\end{align}
where $d_\mathrm{ind}$ is the induced lab-frame dipole moment.
This Hamiltonian describes the first-order dipole-dipole interactions,
but neglects the internal structure of the molecular dipoles.
This approximation is accurate for $R$ large enough that the dipole-dipole interaction is weak compared to the internal structure, \emph{i.e.}, rotational constants.
Where this fails, the interactions are necessarily also strong compared to the kinetic energy available in ultracold collisions.
Deviations from first-order dipolar interactions are therefore expected to occur only where the local momentum is high,
and the flux that reaches such short separations is expected to proceed to short range unaffected by the modification of the interactions,
as long as the interaction does not become repulsive.
Coupling to higher rotational levels, neglected in Eq.~\eqref{eq:BCT}, can only make the lowest adiabatic potential more attractive, not repulsive.
Hence, these deviations from purely first-order dipolar interactions should not affect the short-range loss rate.

\subsection{Microwave electric fields \label{sec:MW}}

\begin{figure*}
\begin{center}
\includegraphics[width=0.425\textwidth]{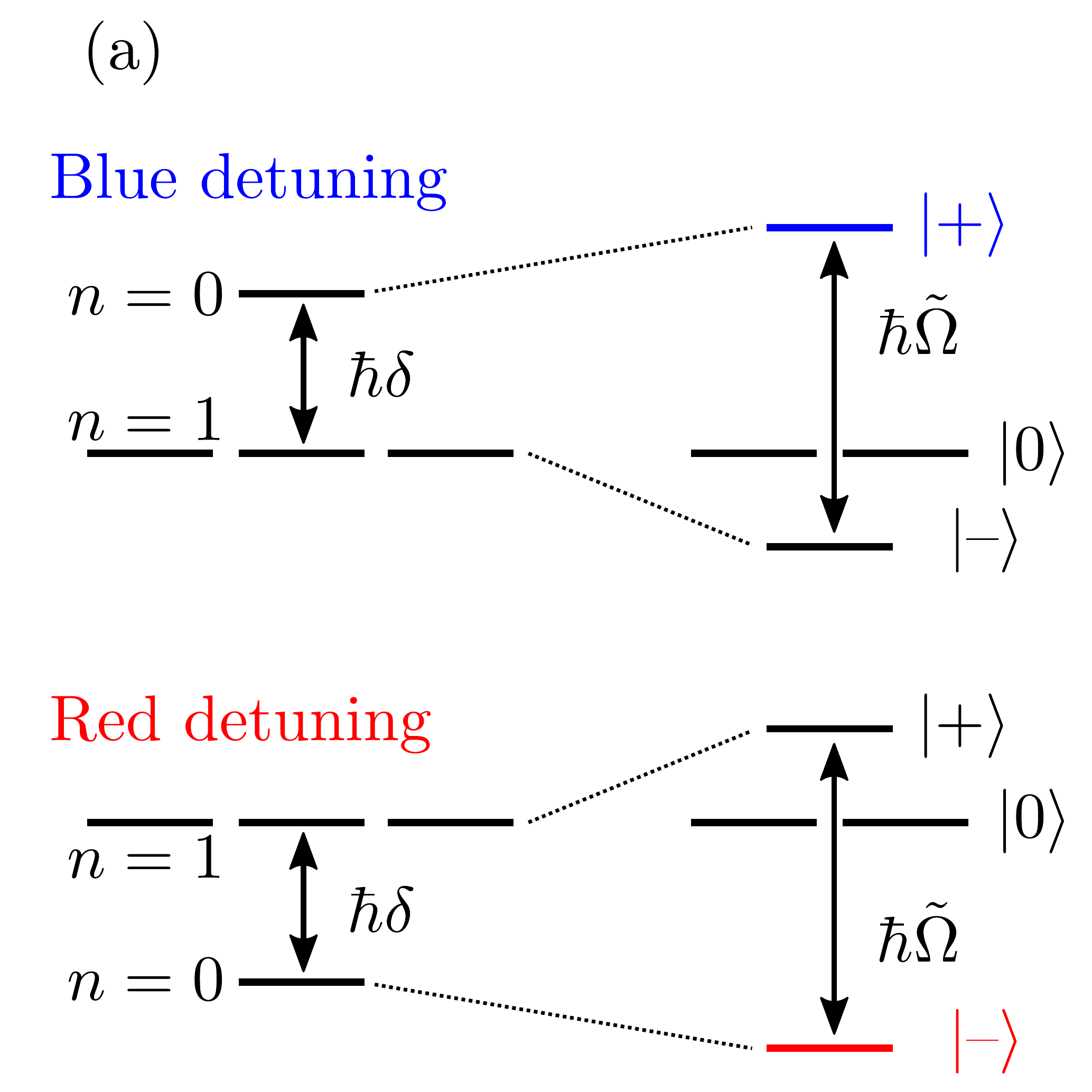}
\includegraphics[width=0.55\textwidth]{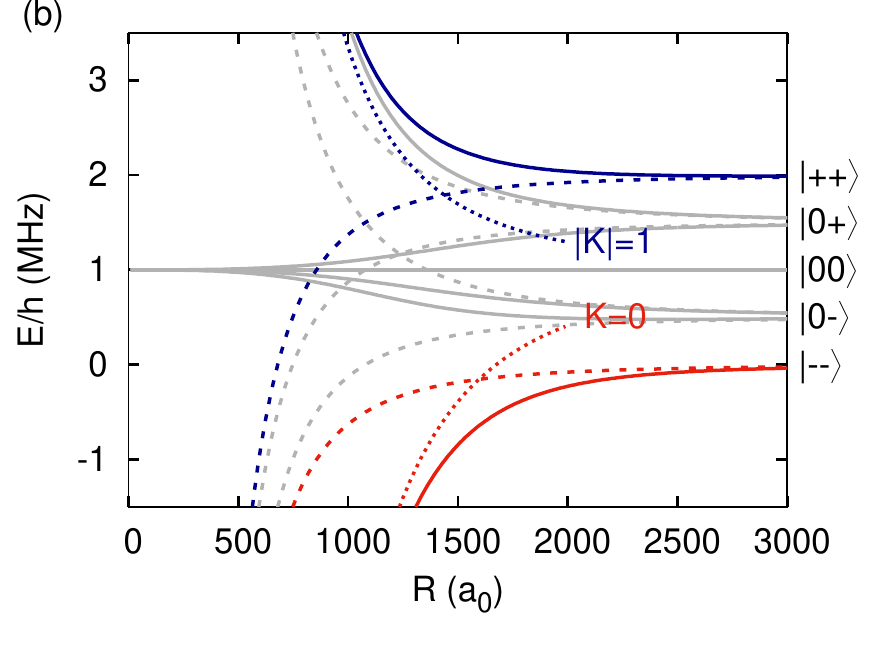}
\caption{ \label{fig:MWschematic}
Level schematic for molecules dressed with microwaves with $\tOmega=2\pi \times 1$~MHz.
(a) Single molecule energy levels for red and blue detuning.
(b) Potential curves for resonant dressing with fixed orientation of the intermolecular axis at $\theta=90^\circ$ relative to the laboratory $z$ axis,
for $\sigma^+$ circular polarization.
Adiabatic potential curves are shown as solid lines,
the first-order dipolar interactions are shown as dashed lines,
and resonant dipole-dipole interactions are shown as dotted lines.
Thresholds used for red and blue detuning are highlighted in color.
}
\end{center}
\end{figure*}

Next, we consider inducing dipole-dipole interactions using microwave radiation tuned close to the lowest ${J=0\rightarrow1}$ rotational transition.
The molecular states are denoted $|J,m_J,N\rangle$ where $J$ is the rotational quantum number, $m_J$ the space-fixed $z$ projection, and $N$ the number of photons relative to a large reference number, $N_0$.
The bare ground state, $|J,m_J,N\rangle=|0,0,0\rangle$, is brought into resonance with the three $m_J=-1,0,1$ bare excited states $|1,m_J,-1\rangle$.
The bare ground state is coupled to a single substate of the excited state, $|1,m,-1\rangle$, where $m=0$ for $\pi$ polarized microwaves and $m=\pm1$ for $\sigma^\pm$ polarized microwaves.
Here, we have defined the space-fixed reference frame relative to the microwave polarization,
\emph{i.e.}, along the direction of linear polarization, or perpendicular to the plane of circular polarization.
For a different but definite microwave polarization,
the coupling is likewise to a single ``bright state'' linear combination of these three bare excited states.
The eigenstates at energies $\pm \frac{1}{2} \tOmega$ are
\begin{align}
|+\rangle &= \cos\varphi |0,0,0\rangle + \sin\varphi |1,m,-1\rangle, \nonumber \\
|-\rangle &= -\sin\varphi |0,0,0\rangle + \cos\varphi |1,m,-1\rangle,
\end{align}
where $\varphi=\mathrm{atan}[(\delta+\sqrt{\delta^2+\Omega^2})/\Omega]$ and $\delta$, $\Omega$, and $\tOmega = \sqrt{\Omega^2+\delta^2}$ are the detuning, Rabi frequency, and generalized Rabi frequency, respectively.
The detuning is defined as $\delta = \omega - 2B_\mathrm{rot}/\hbar$ and $\delta>0$ is referred to as blue detuning and $\delta<0$ as red.
The remaining two excited substates with $m_J\neq m$ are not coupled by the microwave field, and are eigenstates at energy $\hbar\delta$.
These spectator states, not involved in the microwave dressing, are schematically denoted $|0\rangle$.
The level schematics discussed here can be seen in Fig.~\ref{fig:MWschematic}(a).

The field-dressed eigenstates, $|\pm\rangle$, are superpositions of the rotational ground and excited state with different photon numbers,
which can be interpreted as having a dipole moment that oscillates at the microwave drive frequency.
For resonant dressing, $\delta=0$, and linear $z$ polarization, $m=0$,
the dipole moment oscillates along the laboratory $z$ axis, $\bm{d}(t) = {d/\sqrt{3} \cos(\omega t) \hat{\bm{z}}}$.
For circular polarization, $m=1$, the dipole moment rotates in the $xy$ plane,
${\bm{d}(t) = d/\sqrt{3} [\cos(\omega t) \hat{\bm{x}} + \sin(\omega t) \hat{\bm{y}}]}$.
For off-resonant dressing, $\delta\neq 0$, the oscillating dipole moment induced is only a fraction $1/\sqrt{1+(\delta/\Omega)^2}$ of the dipole moment achieved by resonant dressing.

Two oscillating dipoles experience first-order dipole-dipole interactions.
For resonant dressing, the time-averaged first-order interaction is given by
\begin{align}
\langle& ++ | \hat{V} | ++ \rangle = \langle -- | \hat{V} | -- \rangle = \frac{1}{4\pi\epsilon_0 R^3} \nonumber \\
&\times \begin{cases}
- d^2/3 P_2(\cos\theta) & \text{for linear polarization, $m=0$,} \\
+ d^2/6 P_2(\cos\theta) & \text{for circular polarization, $m=|1|$,}
\end{cases}
\label{eq:MWBCT}
\end{align}
where $P_2(z)=(3z^2-1)/2$ is a Legendre polynomial.
That is, for resonant dressing with linearly polarized microwaves,
the first-order interactions induced are identical to dipolar interactions with an effective dipole moment $d_\mathrm{eff} = d/\sqrt{6}$, compare Eqs.~\eqref{eq:BCT} and \eqref{eq:MWBCT}.
For circular polarization, the effective dipole moment is smaller, $d_\mathrm{eff} = d/\sqrt{12}$, and the sign of the interaction reversed.
These interactions correspond precisely to the classical time-averaged interactions between the dipoles oscillating along $z$ or rotating in the $xy$ plane, respectively.
For off-resonant dressing, $\delta\neq 0$, only a fraction of the effective dipole moment is induced,
which depends on the ratio $|\delta|/\Omega$, but not on the generalized Rabi frequency, $\tOmega$.

\begin{figure*}
\begin{center}
\includegraphics[width=0.475\textwidth]{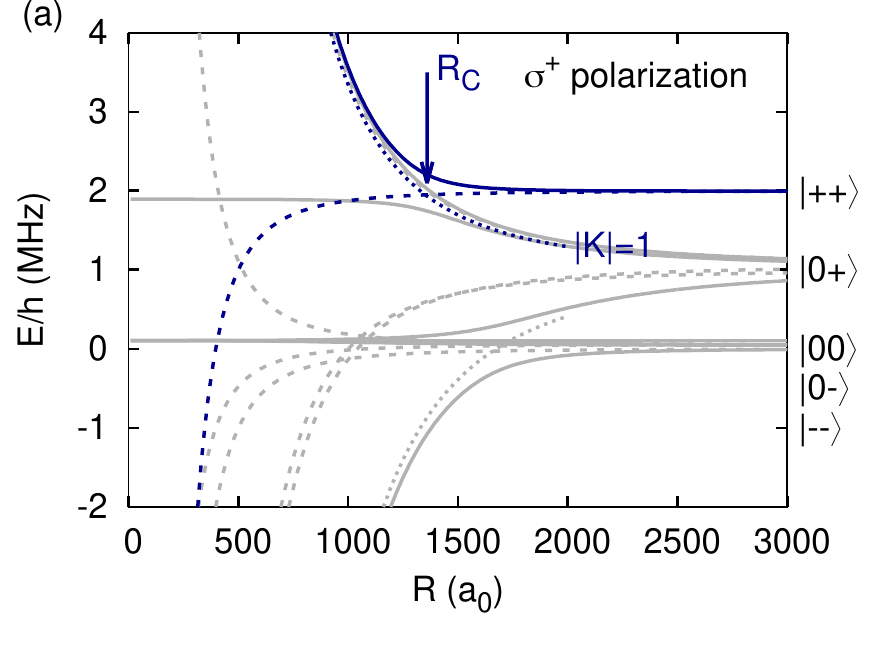}
\includegraphics[width=0.475\textwidth]{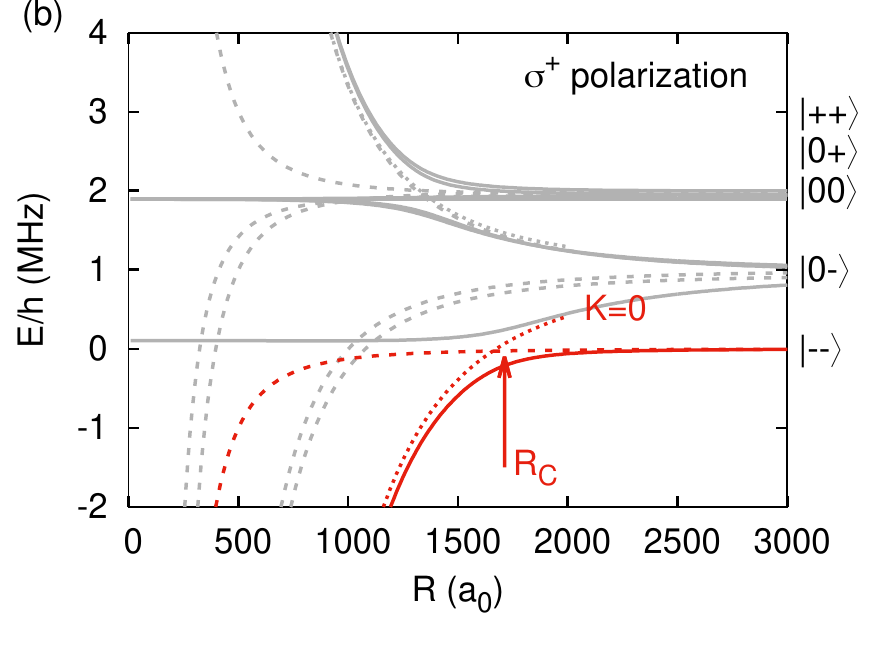}
\includegraphics[width=0.475\textwidth]{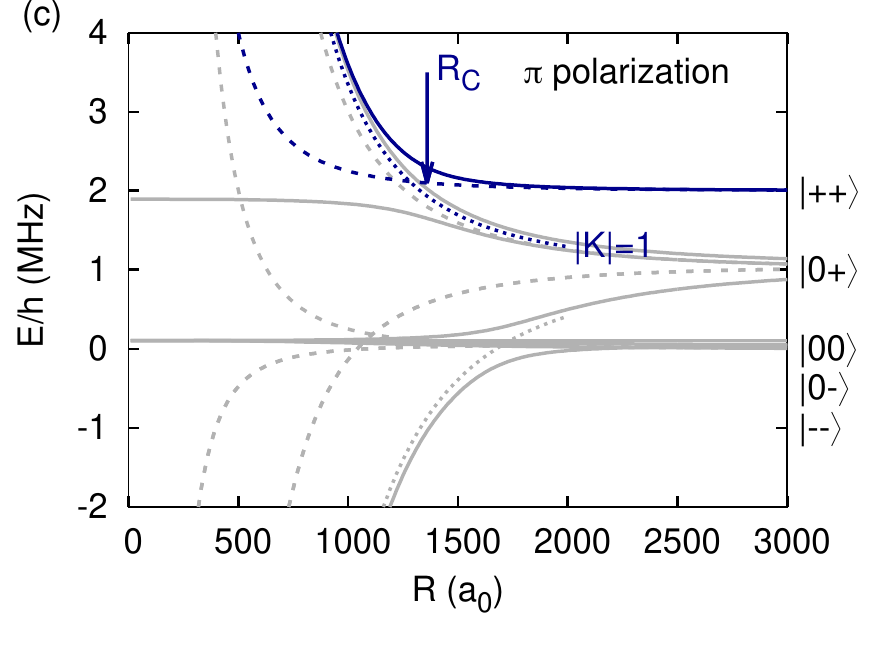}
\includegraphics[width=0.475\textwidth]{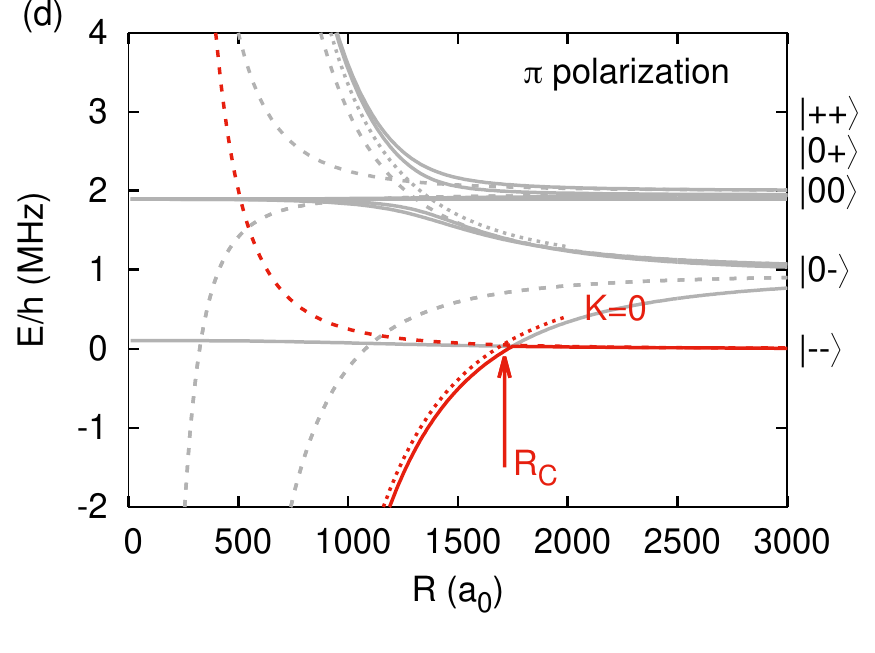}
\caption{ \label{fig:MWoffresonance}
Potential energy curves for off-resonant microwave dressing, $\tOmega=2\pi \times 1$~MHz and $|\delta|/\Omega=3$, and at fixed $\theta=90^\circ$, under the following conditions
(a) Blue detuning, circular polarization,
(b) Red detuning, circular polarization,
(c) Blue detuning, linear polarization,
(d) Red detuning, linear polarization.
The correspondence to molecules in the upper and lower field dressed level is indicated on the right.
For blue detuning, the initial state corresponds to two molecules in the upper field-dressed level, $|+\rangle$,
and for far off-resonant dressing the lower field-dressed level, $|-\rangle$, is nearly degenerate with the spectator states, $|0\rangle$.
For red detuning the initial state corresponds to the lower field-dressed level, $|-\rangle$,
and in the case of off-resonant dressing $|+\rangle$ and $|0\rangle$ are nearly degenerate.
}
\end{center}
\end{figure*}

The perturbative treatment of the dipole-dipole interaction breaks down where this interaction is comparable to the energy spacing of internal states of the molecule.
For a molecule in field-free space or a static electric field, this occurs essentially where the dipole-dipole interaction is comparable to the rotational constant, at very short separation $R\propto (d^2/B_\mathrm{rot})^{1/3}$.
In the microwave-dressed case, however, there exist oscillating transition dipole moments between the various field-dressed levels,
and their spacings are on the order of $\tOmega$, which can be orders of magnitude smaller than the rotational constant.
In this case, dipole-dipole coupling among the nearly-degenerate states can easily become the dominant term in the Hamiltonian.
The molecule-molecule field-dressed eigenstates are $|J_A, M_{JA}\rangle|J_B, M_{JB}\rangle|N\rangle = |0,0\rangle|0,0\rangle|0\rangle$ and $|1,M_J\rangle|1,M_J'\rangle|-2\rangle$, which experience no dipole-dipole interaction,
and the body-fixed states $\hat{\mathcal{R}} |00,1K\rangle_s |-1\rangle$ that do.
Here, the subscript $s$ denotes symmetrization, i.e. $|00,1K\rangle_s = [ |0,0\rangle|1,K\rangle + |1,K\rangle|0,0\rangle ]/\sqrt{2}$.
Here, $\hat{\mathcal{R}}$ is an operator that rotates the laboratory $z$ axis to the intermolecular axis, and $K=-1,0,1$ is the projection of rotational angular momentum onto the intermolecular axis.
The states with $K=\pm1$ experience a repulsive interaction $+d^2/(3 R^{3})$,
whereas the $K=0$ state experiences an attractive interaction $-2d^2/(3 R^{3})$.
This is referred to as resonant dipole-dipole interactions,
as these arise from transition dipole moments for excitation of one molecule from $J=0$ to 1 and simultaneous de-excitation of the other from $J=1$ to 0.

Figure~\ref{fig:MWschematic}(b) shows adiabatic potential curves for resonant dressing, $\delta=0$, with $\sigma^+$ circular polarization and $\tOmega=2\pi \times 1$~MHz.
The curves are obtained by diagonalizing the Hamiltonian for fixed $R$ and orientation of the intermolecular axis, at an angle $\theta=90^\circ$ relative to the laboratory $z$ axis.
Couplings to states outside of the nearly degenerate set shown are neglected.
Potential curves for red and blue detuning are highlighted in color.
The long-range first-order dipole-dipole interaction is shown as the corresponding dashed lines,
and the short-range resonant dipole-dipole interaction as the dotted lines, labeled by $|K|=0$ and $1$ for attractive and repulsive interactions, respectively.

Figure~\ref{fig:MWoffresonance} shows adiabatic potential curves for off-resonant dressing, where $|\delta|/\Omega=3$ and $\tOmega=2\pi \times 1$~MHz.
The off-resonant dressing induces only a small fraction of the maximum oscillating dipole moment,
and resonant dipole-dipole interactions become dominant well before the weak first-order dipole-dipole interaction becomes comparable to $\tOmega\approx|\delta|$.
At very low collision energies, this first-order dipole-dipole interaction could still be dominant and determine the flux of colliding molecules that reach short range.
However, at collision energies high compared to $\tOmega$, only the resonant dipole-dipole interactions are appreciable compared to the kinetic energy,
and completely determine the collision dynamics.
As discussed above, the singly excited threshold is split by resonant dipole-dipole interactions into an attractive $K=0$ and a degenerate repulsive $K=\pm1$ branch.
For red detuning, the attractive $K=0$ branch crosses the flat initial potential from above at the Condon point $R_\mathrm{C}^{(\mathrm{r})}=(2d_0^2/3\tOmega)^{1/3}$.
The crossing between the initial and excited state is avoided by the Rabi coupling.
Because the transition dipole moment is along the intermolecular axis,
the effective Rabi frequency due to the microwaves polarized in the lab frame depends on the frame transformation as the Wigner $d$-matrix element $d^{(1)}_{m,0}(\theta)$,
where $m=0$ for linear and $\pm 1$ for circular polarization.
For blue detuning, the repulsive $K=\pm1$ branch crosses the initial potential from below at the Condon point $R_\mathrm{C}^{(\mathrm{b})}=(d_0^2/3\tOmega)^{1/3}$.
The effective Rabi coupling to the $K=\pm 1$ states then depends on the orientation of the intermolecular axis, at polar angles $\theta$ and $\phi$, as $D^{(1)}_{m,\pm 1}(\theta,\phi)$.
One notable feature is that for blue detuning and circular polarization ($m=\pm 1$),
the Rabi coupling to the bright-state combination of $K=\pm 1$ does not vanish for any orientation of the intermolecular axis, unlike for linear polarization ($m=0$) or red detuning ($K=0$)~\cite{karman:18d,karman:19c}.

We note that the present discussion of interaction potentials for fixed orientation of the intermolecular axis is a simplification made to analyze the interaction potentials only.
In reality the scattering wavefunction does not localize for some sharply defined orientation,
and one must quantum mechanically average over angles according to the wavefunction for $s$ or $p$-wave collisions, respectively.
The anisotropic dipole-dipole interactions lead to some orientation, fully accounted for by the mixing of different partial waves in the coupled-channels calculations outlined in Sec.~\ref{sec:theory}.

\begin{widetext}
\subsection{Full description of MW-induced interactions}

Above, we have described first-order dipole-dipole interactions induced by microwave dressing, for large $\tOmega$.
as well as the case of resonant dipole-dipole interactions for perturbatively small $\Omega$.
Here, we consider a more complete case while simplifying the analysis by assuming a fixed orientation of the intermolecular axis, $\hat{R}$, at an angle $\theta$ with respect to the laboratory $z$ axis,
the frame in which the microwave polarization is defined.
We consider linear $\pi$ polarization here,
and refer the reader to Ref.~\cite{yan:20} for a similar analysis in the case of $\sigma^+$ polarization.
We consider states for a pair of colliding molecules and the microwave field.
The bare ground state is denoted as $|J_A M_A, J_B M_B\rangle|N-N_0\rangle = |00,00\rangle|0\rangle$.
In addition we consider the following nearly degenerate states,
$|00,1\, -1\rangle_s|-1\rangle$, $|00,1\, 0\rangle_s|-1\rangle$, and $|00,1\, 1\rangle_s|-1\rangle$,
at energy $-\hbar\delta$,
and
$|1\, -1,1\, -1\rangle_s|-2\rangle$,
$|1\, -1,1\,  0\rangle_s|-2\rangle$, 
$|1\, -1,1\,  1\rangle_s|-2\rangle$, 
$|1\,  0,1\,  0\rangle_s|-2\rangle$,
$|1\,  0,1\,  1\rangle_s|-2\rangle$,
and $|1\, 1,1\,  1\rangle_s |-2\rangle$,
at energy $-2\hbar\delta$.
The subscript $s$ denotes symmetrization with respect to exchange of the two molecules.
We note that the ``primitive'' basis functions used here are not products of eigenstates of the non-interacting molecules in the microwave field.
Instead, a $\pi$-polarized microwave field couples $|00\rangle|N\rangle$ and $|10\rangle|N-1\rangle$ at a strength parameterized by the Rabi frequency, $\Omega$.
Dipole-dipole interactions occur between states with the same microwave photon number and opposite parity for both molecular states.
Hence, within the nearly degenerate manifold, dipole-dipole coupling occurs only between the singly excited states.
Together, this leads to the following matrix representation of the Hamiltonian
\begin{align}
\bm{H} = 
\begin{pmatrix}
 0        & 0        & \hbar\Omega/\sqrt{2} & 0        & 0        & 0        & 0        & 0        & 0        & 0        \\ 
 0        & \frac{1}{3} \frac{d_0^2}{R^{3}} C_{2,0}(\hat{R}) -\hbar\delta        & \frac{1}{\sqrt{3}} \frac{d_0^2}{R^{3}} C_{2,1}(\hat{R})        & \sqrt{\frac{2}{3}} \frac{d_0^2}{R^{3}} C_{2,2}(\hat{R})        & 0               & \hbar\Omega/2 & 0        & 0        & 0        & 0        \\
 \hbar\Omega/\sqrt{2}        & -\frac{1}{\sqrt{3}} \frac{d_0^2}{R^{3}} C_{2,-1}(\hat{R})        & -\frac{2}{3} \frac{d_0^2}{R^{3}} C_{2,0}(\hat{R})-\hbar\delta        & -\frac{1}{\sqrt{3}} \frac{d_0^2}{R^{3}} C_{2,1}(\hat{R})        & 0        & 0        & 0        & \hbar\Omega/\sqrt{2} & 0        & 0        \\
 0        & \sqrt{\frac{2}{3}} \frac{d_0^2}{R^{3}} C_{2,-2}(\hat{R})        & \frac{1}{\sqrt{3}} \frac{d_0^2}{R^{3}} C_{2,-1}(\hat{R})        & \frac{1}{3} \frac{d_0^2}{R^{3}} C_{2,0}(\hat{R})-\hbar\delta        & 0        & 0        & 0        & 0        &  \hbar\Omega/2   & 0     \\
 0        & 0        & 0        & 0        & -2\hbar\delta        & 0        & 0        & 0        & 0        & 0        \\
 0        & \hbar\Omega/2        & 0        & 0        & 0        & -2\hbar\delta        & 0        & 0        & 0        & 0        \\
 0        & 0        & 0        & 0        & 0        & 0        & -2\hbar\delta        & 0        & 0        & 0        \\
 0        & 0        & \hbar\Omega/\sqrt{2}        & 0        & 0        & 0        & 0        & -2\hbar\delta        & 0        & 0        \\
 0        & 0        & 0        & \hbar\Omega/2        & 0        & 0        & 0        & 0        & -2\hbar\delta        & 0        \\
 0        & 0        & 0        & 0        & 0        & 0        & 0        & 0        & 0        & -2\hbar\delta        
\end{pmatrix}.
\end{align}
Because there are no dipole-dipole interactions among the doubly excited states,
and the $\pi$-polarized microwave field couples the ground state only to $m=0$ excited states, doubly excited states with both $m_A$ and $m_B \neq 0$ are completely uncoupled and can be removed from the basis.
This results in
\begin{align}
\bm{H} =
\begin{pmatrix}
 0        & 0        & \hbar\Omega/\sqrt{2} & 0        & 0                & 0        & 0                    \\ 
 0        & \frac{1}{3} \frac{d_0^2}{R^{3}} C_{2,0}(\hat{R}) -\hbar\delta        & \frac{1}{\sqrt{3}} \frac{d_0^2}{R^{3}} C_{2,1}(\hat{R})        & \sqrt{\frac{2}{3}} \frac{d_0^2}{R^{3}} C_{2,2}(\hat{R})                      & \hbar\Omega/2         & 0        & 0                \\
 \hbar\Omega/\sqrt{2}        & -\frac{1}{\sqrt{3}} \frac{d_0^2}{R^{3}} C_{2,-1}(\hat{R})        & -\frac{2}{3} \frac{d_0^2}{R^{3}} C_{2,0}(\hat{R})-\hbar\delta        & -\frac{1}{\sqrt{3}} \frac{d_0^2}{R^{3}} C_{2,1}(\hat{R})                & 0                & \hbar\Omega/\sqrt{2} & 0               \\
 0        & \sqrt{\frac{2}{3}} \frac{d_0^2}{R^{3}} C_{2,-2}(\hat{R})        & \frac{1}{\sqrt{3}} \frac{d_0^2}{R^{3}} C_{2,-1}(\hat{R})        & \frac{1}{3} \frac{d_0^2}{R^{3}} C_{2,0}(\hat{R})-\hbar\delta               & 0                & 0        &  \hbar\Omega/2       \\
 0        & \hbar\Omega/2        & 0        & 0               & -2\hbar\delta               & 0               & 0        \\
 0        & 0        & \hbar\Omega/\sqrt{2}               & 0                & 0        & -2\hbar\delta               & 0        \\
 0        & 0        & 0        & \hbar\Omega/2           & 0            & 0        & -2\hbar\delta              \\
\end{pmatrix}
\end{align}
in the basis
$\{ |00,00\rangle, |00,1\,-1\rangle_s, |00,1\,0\rangle_s, |00,1\,+1\rangle_s, |1\, -1,1\,  0\rangle_s, |1\,  0,1\,  0\rangle_s, |1\, 0,1\,  1\rangle_s \}$
where the microwave photon number is implicit. 

For large detuning, $|\delta|\gg\Omega$, the Rabi coupling can be treated perturbatively,
and plays a role only at the Condon point, where two potential curves cross as the resonant dipole-dipole interaction compensates for the detuning.
Therefore, the doubly excited channels can be ignored, leading to a $4\times 4$ problem.
The resonant dipole-dipole interaction is most conveniently described in the body-fixed frame,
which has the $z$ axis along the intermolecular frame.
Body-referred states with projection of angular momentum onto the intermolecular axis $K$ are given by
\begin{align}
\hat{\mathcal{R}} |00,1K\rangle_s = \hat{\mathcal{R}} \left[ |00,1K\rangle + |1K,00\rangle \right]/\sqrt{2}.
\end{align}
These are eigenstates of the dipole-dipole interaction
\begin{align}
\hat{V} \hat{\mathcal{R}} |00,1K\rangle_s = \hat{\mathcal{R}} |00,1K\rangle_s \times \begin{cases} +\frac{d^2}{3R^3}  & \text{for $K=\pm1$,} \\ -\frac{2d^2}{3R^3}  & \text{for $K=0$.} \end{cases}
\end{align}
In the body-fixed frame,
the effective Rabi coupling depends on the orientation of the intermolecular axis with respect to the lab frame,
in which the microwave polarization is defined,
$\sigma=0,1$ for $\pi$ or $\sigma^+$ polarization,
\begin{align}
\hat{\mathcal{R}}^\dagger \hat{d}_\sigma \hat{\mathcal{R}} = \sum_{\kappa} D^{(1)\ast}_{\sigma,\kappa}(\hat{R}) \hat{d}_{\kappa}.
\end{align}
This leads to the Hamiltonian
\begin{align}
\bm{H} =
\begin{pmatrix}
 0                           & \hbar\Omega\sqrt{2} D^{(1)\ast}_{0,+1}(\hat{R}) & \hbar\Omega\sqrt{2} D^{(1)\ast}_{0,0}(\hat{R})  &  \hbar\Omega\sqrt{2} D^{(1)\ast}_{0,-1}(\hat{R}) \\
\hbar\Omega\sqrt{2} D^{(1)}_{0, 1}(\hat{R}) & \frac{d_0^2}{3R^3}-\hbar\delta  & 0                                     &  0 \\
\hbar\Omega\sqrt{2} D^{(1)}_{0, 0}(\hat{R}) & 0                                  & -\frac{2d_0^2}{3R^3}-\hbar\delta   &  0 \\
\hbar\Omega\sqrt{2} D^{(1)}_{0,-1}(\hat{R}) & 0                                  & 0                                     &  \frac{d_0^2}{3R^3}-\hbar\delta 
\end{pmatrix}
\end{align}
For red detuning, $\delta>0$, the attractive resonant dipole-dipole interaction for $K=0$ causes a crossing with the ground-state potential at the Condon point $R_\mathrm{C}^{(\mathrm{r})} = (-2d_0^2/3\hbar\delta)^{1/3}$.
The effective Rabi coupling then depends on the angle $\theta$ between linear $\pi$ microwave polarization and the intermolecular axis as $D^{(1)\ast}_{0,0}(\hat{R})=\cos\theta$.
For blue detuning, $\delta<0$, the situation is reversed and the repulsive resonant dipole-dipole interaction for $K=\pm 1$ causes a crossing at the Condon point $R_\mathrm{C}^{(\mathrm{b})}=(d_0^2/3\tOmega)^{1/3}$.
In this case, the Rabi coupling is to the bright-state combination of $K=\pm 1$, and its magnitude depends on the angle $\theta$ between the orientation of the intermolecular axis and the linear polarization direction as $\sin\theta$.
The resulting potential curves and crossing at the Condon point avoided by the Rabi frequency are sketched in Fig.~\ref{fig:cartoon}.

A similar discussion for $\sigma^+$ polarized microwaves can be found in the supplement of Ref.~\cite{yan:20}.

\end{widetext}

\section{Coupled-channels calculations \label{sec:theory}}

Using the tools for tuning interactions using static fields and microwave dressing developed in Sec.~\ref{sec:dip},
we are now ready to treat molecular collisions within the coupled-channels framework.
The two molecules, $X=A,B$, are modeled as rigid rotors with a dipole moment,
\begin{align}
\hat{H}^{(X)} = B_\mathrm{rot} \hat{J}^{(X)\,2}- \hat{\bm{d}}^{(X)} \cdot \efield + \hat{H}_\mathrm{ac}^{(X)},
\label{eq:hmon}
\end{align}
with rotational constant $B_\mathrm{rot}$.
The second term represents the Stark interaction with the static field.
The last term represents the interaction with a microwave field~\cite{cohentannoudji:98},
\begin{align}
\hat{H}_\mathrm{ac}^{(X)} = -\sqrt{\frac{\hbar\omega}{2\epsilon_0 V_0}}
\left[ \hat{d}_\sigma^{(X)} \hat{a}_\sigma +
\hat{d}_\sigma^{(X)\dagger} \hat{a}_\sigma^\dagger\right]
        + \hbar\omega  \left(\hat{a}_\sigma^\dagger \hat{a}_\sigma - N_0\right).
\label{eq:Hac}
\end{align}
Here, $N_0 = \epsilon_0 E_\mathrm{ac}^2 V_0 /2\hbar\omega$ is the reference
number of photons in a reference volume, $V_0$, at microwave electric field
strength, $E_\mathrm{ac}$~\cite{avdeenkov:15}.
The operators $\hat{a}_\sigma^\dagger$ and $\hat{a}_\sigma$ are creation and annihilation operators for photons at angular frequency $\omega$ in polarization mode $\sigma$.
For linear polarization, $\sigma=0$, whereas for circular polarization, $\sigma=\pm 1$,
and $\hat{d}_0 = \hat{d}_z$ and $\hat{d}_{\pm 1} = \mp (\hat{d}_x\pm i \hat{d}_y)/\sqrt{2}$ are the usual spherical components of the dipole operator.

The total Hamiltonian is
\begin{align}
\hat{H} = -\frac{\hbar^2}{2 \mu} \frac{d^2}{dR^2} 
+ \frac{\hbar^2 \hat{L}^2}{2 \mu R^2} + \hat{H}^{(A)} + \hat{H}^{(B)} + \hat{V}(R).
\label{eq:Htot}
\end{align}
Here $\mu$ is the reduced mass, $R$ is the intermolecular distance
and $\hat{L}$ is the angular momentum operator associated with the end-over-end rotation. The first and second terms describe radial and centrifugal kinetic energy, respectively.
The final term is the dipole-dipole interaction between the molecules~\cite{karman:18d}.
This accounts for first-order dipole-dipole interactions if the molecules are polarized,
as well as for the second-order dipole-dipole interaction -- so-called rotational van der Waals interaction -- which is dominant in the absence of external fields.
The molecular parameters chosen here,
$B_\mathrm{rot}/h = 2.8217$~GHz and $d = 2.72$~Debye,
correspond to NaK molecules.
Hyperfine interactions are not included.
For NaK, and many bi-alkali molecules such as RbCs, NaRb, and KCs, the rotational contribution to the van der Waals interaction is far dominant over the electronic contribution,
which is neglected here,
but we note this can be important for molecules with smaller dipole moments or larger rotational constants, such as KRb.

We use the basis functions
\begin{align}
|J_A M_A\rangle|J_B M_B\rangle|L M_L\rangle|N\rangle,
\end{align}
where $|J_A M_A\rangle$ ($|J_B M_B\rangle$) describes the rotation of molecule $A$ ($B$),
$|L M_L\rangle$ describes the end-over-end rotation of the collision complex,
and $|N\rangle$ number of microwave photons.
For bosons (fermions), only even (odd) $L$ are included and the basis functions are adapted to permutation symmetry~\cite{karman:18d}.
The calculations are converged to several percent accuracy while truncating the basis at $L_\mathrm{max}=6$,
$J_\mathrm{max}=1$ and $N-N_0 = -2, -1, 0$ for calculations with a microwave electric field,
and $J_\mathrm{max}=3$ and $N-N_0 = 0$ for calculations with a static field.
Calculations are performed separately for different values of the conserved $z$-projection of total angular momentum 
\begin{align}
\mathcal{M} = M_A + M_B + M_L + \sigma N,
\end{align}
where the electric field is along the $z$ direction,
and the microwave-polarization is defined as described above.

Coupled-channels calculations use the renormalized Numerov algorithm~\cite{johnson:78} with absorbing boundary conditions at short range~\cite{janssen:13}.
The absorbing boundary condition is imposed at $R_\mathrm{min}=50$~$a_0$,
but the precise loss rates are independent of this choice~\cite{karman:18d}.
Matching to the boundary conditions yields $S$-matrices for combined short-range and long-range loss channels.
The corresponding cross sections are obtained from the T-matrix $\bm{T}=\bm{1}-\bm{S}$ for inelastic scattering from state $i$ to $f$ at long-range as
\begin{align}
\sigma_{i\rightarrow f}^{(\mathrm{inel})}(E) = \frac{2\pi}{k^2} \sum_{L,M_L,L',M_L'} \left| T^{(\mathrm{LR})}_{f,L',M_L';i,L,M_L} \right|^2,
\end{align}
where $k^2=2\mu E$ is the wavenumber, $E$ is the collision energy,
and for short-range capture in adiabatic short-range channel $r$ as
\begin{align}
\sigma^{(\mathrm{SR})}(E) = \frac{2\pi}{k^2} \sum_{r,L,M_L} \left| T^{(\mathrm{SR})}_{r;i,L,M_L} \right|^2.
\end{align}
We use a logarithmically spaced grid of 13 energies between 1~nK and 10~$\mu$K,
and numerically compute loss rates as
\begin{align}
\beta = \sqrt{\frac{8 k_{\rm B}T}{\pi \mu}} \frac{1}{(k_{\rm B}T)^2} \int_0^\infty \sigma(E) \exp\left(-\frac{E}{k_\mathrm{B}T}\right) E dE,
\end{align}
for $T=500$~nK.

The ``space-fixed dipole'' approximation, see Eq.~\eqref{eq:BCT} and the discussion below, permits a much simpler separate implementation.
However, it is here implemented pragmatically by truncating the channel basis prior to propagation to include only the lowest eigenstate of $\hat{H}^{(X)}$ for each molecule.

\section{Loss rates \label{sec:loss}}

\begin{figure}
\begin{center}
\includegraphics[width=0.475\textwidth]{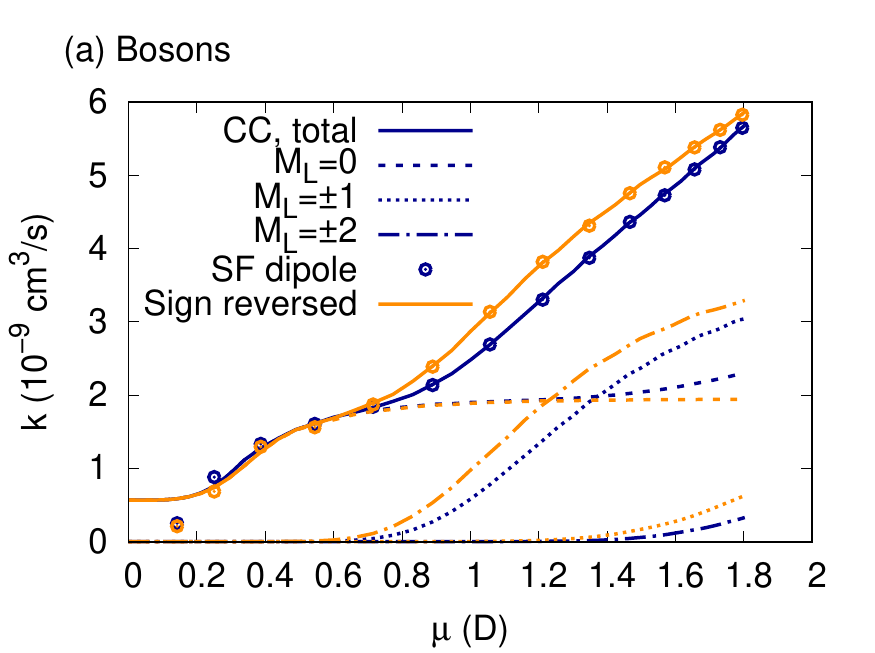}
\includegraphics[width=0.475\textwidth]{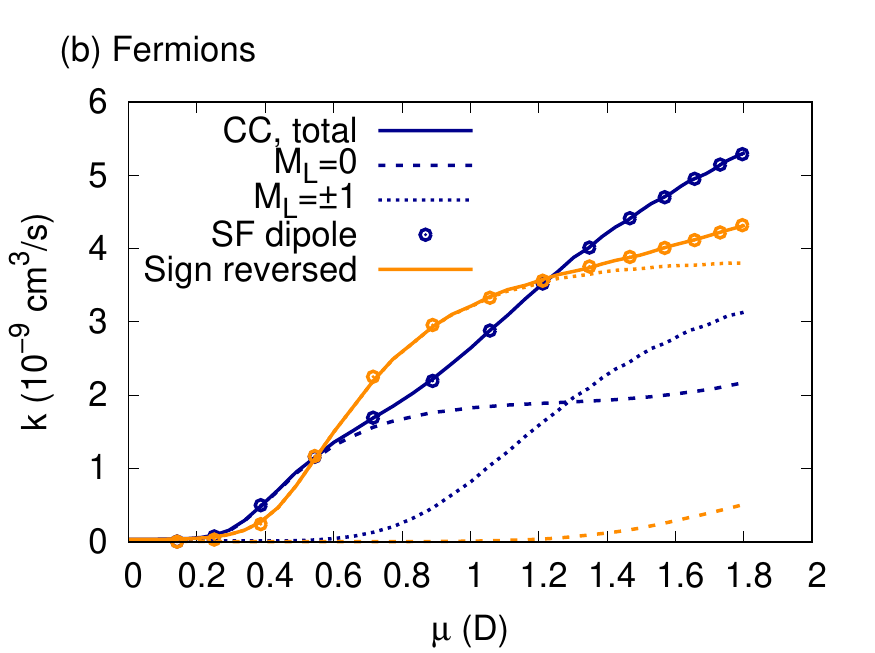}
\caption{ \label{fig:dc}
Loss rates for collisions of (a) bosonic and (b) fermionic NaK molecules polarized by a static electric field (lines).
Dashed lines show the contributions of different $M_L$.
Loss rates for the space-fixed dipole approximation, Eq.~\eqref{eq:BCT}, are shown as markers.
Results shown in blue are shown for the physical dipole-dipole interaction,
whereas results shown in orange have been obtained with the sign of the interaction reversed.
}
\end{center}
\end{figure}

Figure~\ref{fig:dc} shows loss rates for molecules polarized in static electric fields for bosonic [panel(a)] and fermionic [panel(b)] NaK molecules at a temperature of 500~nK.
Results from coupled-channels (c.c.) calculations are shown as solid lines,
which are in close agreement with the space-fixed dipole approximation, Eq.~\eqref{eq:BCT}, shown as the markers.
This figure also includes results obtained with the sign of the dipole-dipole interaction reversed;
for molecules in identical internal states, this cannot be realized physically using static fields,
but it can be realized using circularly polarized microwave fields, to which we will compare below.
For bosonic molecules, $s$-wave collisions ($L=M_L=0$) lead to loss at zero induced dipole moment given by the universal loss rate $7\times10^{-10}$~cm$^{-3}$/s.
With increasing dipole moment, the $s$-wave loss approaches the unitarity limit $1.8\times10^{-9}$~cm$^{-3}$/s.
At larger induced dipole, $L=2$ partial waves start to contribute,
which have attractive first-order interactions for $M_L=\pm1$ that contribute to the loss.
When the sign of the interaction is reversed, the losses behave similarly, except that first-order dipole-dipole interactions are now attractive for $L=2$, $M_L=\pm2$.
In the fermionic case, losses are small for zero dipole moment,
as the molecules are protected from collisions by the $p$-wave centrifugal barrier.
For finite dipole moment, attractive dipole-dipole interactions for $L=1$, $M_L=0$ reduce the barrier height and induce loss that saturates at the unitarity limit,
until coupling to higher $L$ becomes important.
If the sign of the dipole-dipole interaction is reversed,
first-order dipole-dipole interactions are instead attractive for $M_L=\pm1$, which leads to loss that plateaus near twice the unitarity limit.

\begin{figure}
\begin{center}
\includegraphics[width=0.475\textwidth]{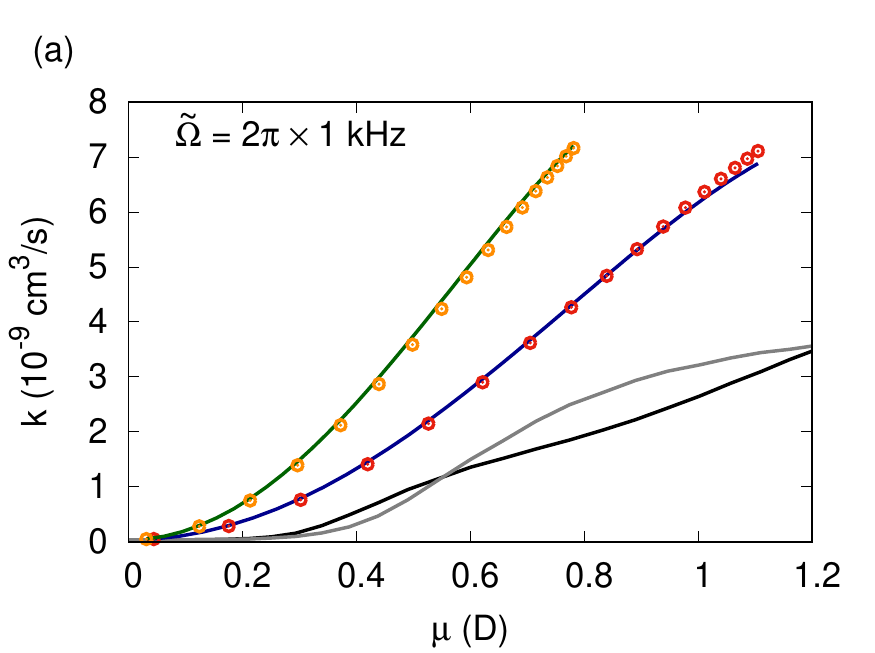}
\includegraphics[width=0.475\textwidth]{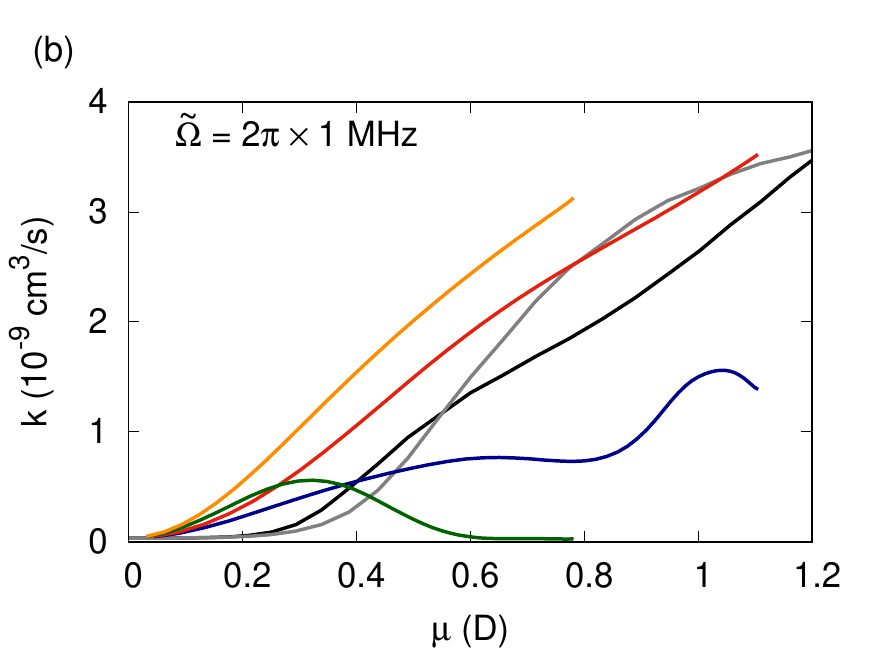}
\includegraphics[width=0.475\textwidth]{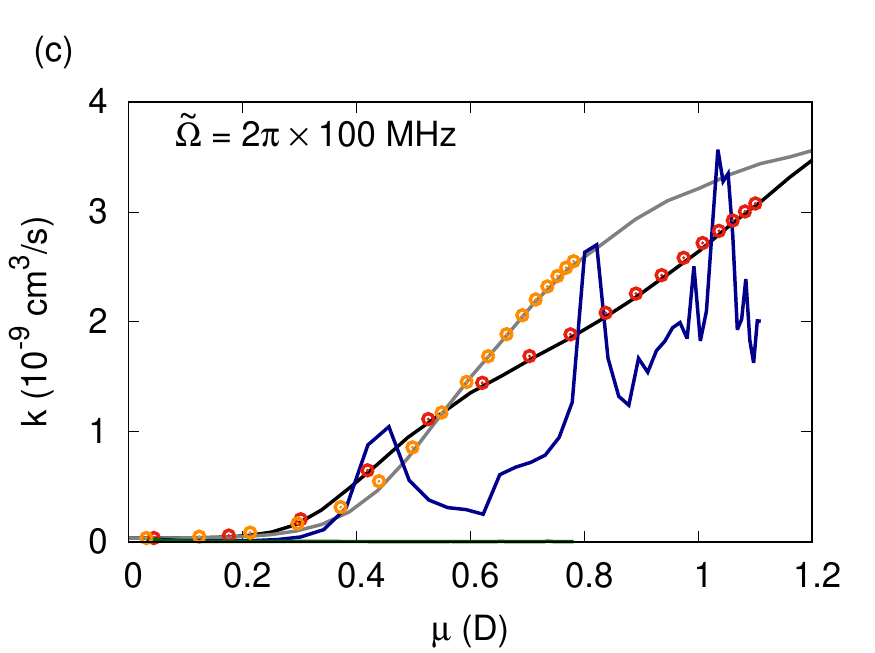}
\caption{ \label{fig:fermion_compare}
Loss rates for fermionic NaK molecules polarized by microwave fields with (a) $\tOmega=2\pi \times 1$~kHz, (b) $\tOmega=2\pi \times 1$~MHz, (c) $\tOmega=2\pi \times 100$~MHz.
Red and orange lines show results at red detuning for linear and circular polarization, respectively.
Blue and green lines show results at blue detuning for linear and circular polarization, respectively.
For comparison, this graph includes the static field loss rates for the physical (sign-reversed) dipole-dipole interaction in black (gray).
For clarity, some of the nearly overlapping lines are instead rendered as markers adhering to the same color coding.
Sharp structure in the observable loss rate stems from stable scattering resonances supported by the long-range interactions, protected from universal short-range loss by repulsive resonant dipolar interactions, See Fig.~\ref{fig:cartoon}(c).
}
\end{center}
\end{figure}

Figure~\ref{fig:fermion_compare} compares loss rates of fermionic NaK molecules as a function of equivalent dipole moment for dipoles induced by static electric fields, and microwave electric fields with red or blue detuning and linear or circular polarization.
Panels (a), (b), and (c) show results obtained using $\tOmega=2\pi \times 1$~kHz, $\tOmega=2\pi \times 1$~MHz, and $\tOmega=2\pi \times 100$~MHz, respectively.
For comparison, this includes loss rates from c.c. calculations for static fields,
which are indistinguishable from the space-fixed dipole approximation.
For interactions induced by dressing with microwave fields,
the space-fixed dipole approximation is approached only for the highest generalized Rabi frequencies and using red detuning.
Using circular polarization, the sign of the dipole-dipole interaction is reversed,
and the resulting loss curve approached at high Rabi frequency is different from that accessible using linear polarization or static fields.
At $\tOmega=2\pi \times 1$~MHz, which is high but achievable experimentally without purpose-built microwave cavities,
the qualitative differences between loss curves obtained using static fields and red-detuned microwave fields are still visible.
Thus, for all realistic Rabi couplings, interactions induced by microwave dressing are to be considered resonant interactions, strongly deviating from expectations based only on the value of the dressed electric dipole moment (the first-order result).

Loss rates obtained using blue-detuned microwaves show qualitatively different behavior.
Rather than approaching the space-fixed dipole approximation for high $\tOmega$,
the losses for circular polarization decrease when dressed on resonance, \emph{i.e.}, for large equivalent dipole moment.
This microwave shielding~\cite{karman:18d} results from repulsive resonant dipole-dipole interactions that occur at shorter range, where the dipole-dipole interaction becomes stronger than $\hbar\tOmega$.
For \emph{linearly} polarized blue-detuned microwaves,
the Rabi coupling to the repulsive branch of the resonant dipolar interaction vanishes for collisions along the polarization direction,
see Sec.~\ref{sec:MW}.
As a result, shielding is ineffective and losses persist at high $\tOmega$ for linear polarization.
Although this leads to ``leaking'' to short range and hence losses do occur,
quasi-stable states exist at long range outside the region of repulsive resonant dipole-dipole interactions,
as illustrated in Fig.~\ref{fig:cartoon}(c).
These long-range bound states lead to resonances, as discussed in Ref.~\cite{lassabliere:18}, that is observable as structured loss rates as a function of induced dipole moment,
which tunes the strength of the long-range interaction that supports these bound states.
For red-detuned microwaves, the resonant dipolar interactions are attractive such that long-range bound states are not protected from universal short-range loss,
and hence these interactions do not support scattering resonances.

Deviations from purely first-order dipolar interactions occur where the dipole-dipole interaction becomes comparable to $\hbar\tOmega$,
and hence these move to ever shorter range for increasing $\tOmega$.
For red-detuned microwaves, the deviations from first-order dipolar interactions are attractive.
If $\tOmega$ is high enough, these deviations occur either in regions that cannot be reached due to repulsive first-order dipole-dipole interaction,
or where the first-order dipole-dipole interaction is already strongly attractive such that it plays no role in reflecting flux back to long range,
and hence does not affect the observable losses.
Therefore, the loss curves for red-detuned microwaves approach the space-fixed dipole approximation for high enough $\tOmega$.
For blue detuning, the deviations from first-order dipolar interactions are repulsive,
and hence can completely transform the interactions in regions otherwise accessible due to attractive first-order dipole-dipole interactions.
Therefore, the loss curves for molecules dressed with blue-detuned microwaves do not approach the space-fixed dipole approximation even for high $\tOmega$.
Again, at realistically achievable $\tOmega$,
resonant dipolar interactions are observed for all microwave polarizations and detunings.

\begin{figure}
\begin{center}
\includegraphics[width=0.475\textwidth]{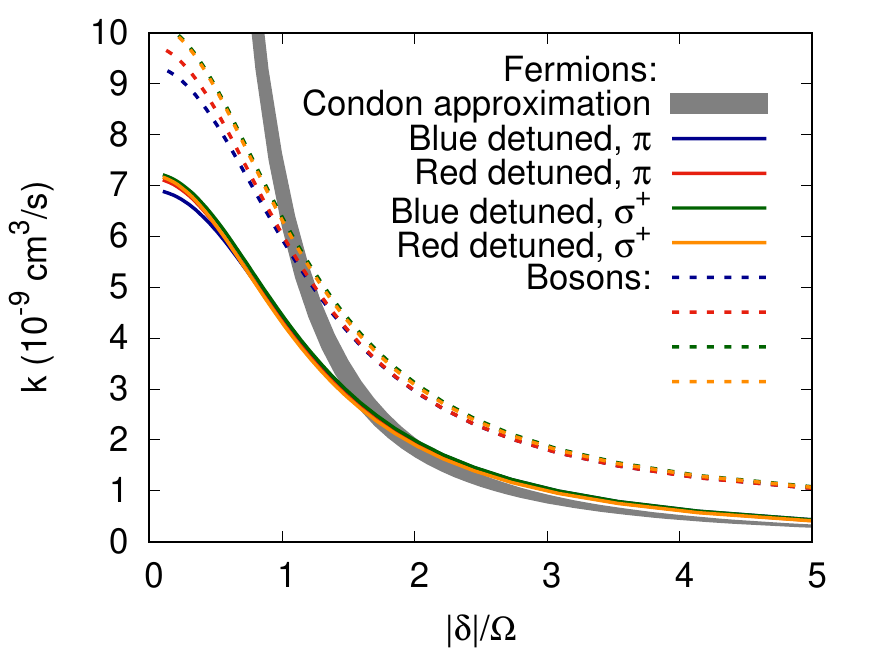}
\caption{ \label{fig:compare_lowest}
Loss rates for fermionic (bosonic) molecules are shown as solid (dashed) lines for $\tOmega=2\pi \times 1$~kHz, from c.c. calculations.
At this Rabi frequency, $\hbar\Omega\ll k_BT$ at 500~nK, such that the interaction is determined solely by the resonant dipole-dipole interactions,
not the space-fixed induced dipole moment.
Hence, loss curves for all polarizations and signs of the detuning collapse onto one another as a function of $\delta/\Omega$.
The Condon approximation is shown as the gray shaded area,
indicating the effect of the difference in Condon point for red and blue detuning.
The Condon approximation is accurate for large detuning,
but misses the background universal loss rate which causes an offset for bosonic molecules.
}
\end{center}
\end{figure}

For the lowest Rabi frequencies $\tOmega =2\pi \times 1$~kHz~$\ll k_BT/\hbar$ at 500~nK, the interaction is dominated by resonant dipole-dipole interactions wherever it is appreciable compared to the kinetic energy,
and parameterization of the interaction in terms of the induced space-fixed oscillating dipole moment, shown on the horizontal axis in Fig.~\ref{fig:fermion_compare}(a), is not the most meaningful.
As shown in Fig.~\ref{fig:MWoffresonance}, the relevant initial state potentials are essentially flat,
but exhibit avoided crossings with attractive or repulsive resonant dipole-dipole potentials, depending on the sign of the detuning.
Losses due to the small Rabi coupling for off-resonant dressing can be described as occurring exclusively at the Condon point,
leading to an accurate analytical expression~\cite{julienne:96,burnett:96,boisseau:00} for the loss rate
\begin{align}
\beta_\mathrm{C} = \frac{16\pi^2}{9\hbar}\, \frac{\Omega^2}{\delta^2} \, d_0^2\, \langle j_l(k R_\mathrm{C})^2 \rangle 
\label{eq:condon}
\end{align}
as shown in Ref.~\cite{yan:20} for red detuning.
We note that nonadiabatic transitions at the Condon point cannot be described by the Landau-Zener formula because the de Broglie wave length is much longer than the region of the crossing.
For blue detuning, Eq.~\eqref{eq:condon} is still valid but the Condon point $R_\mathrm{C}$ in the argument of the spherical Bessel function, $j_l$, is smaller than in the case of red detuning, see Sec.~\ref{sec:MW}.
Hence, the ratio $|\delta|/\Omega$ controls the fraction of the flux that scatters through resonant dipole-dipole interactions,
essentially independently of the sign of the detuning or the microwave polarization.
This can be seen in Fig.~\ref{fig:compare_lowest}.

\begin{figure}
\begin{center}
\includegraphics[width=0.475\textwidth]{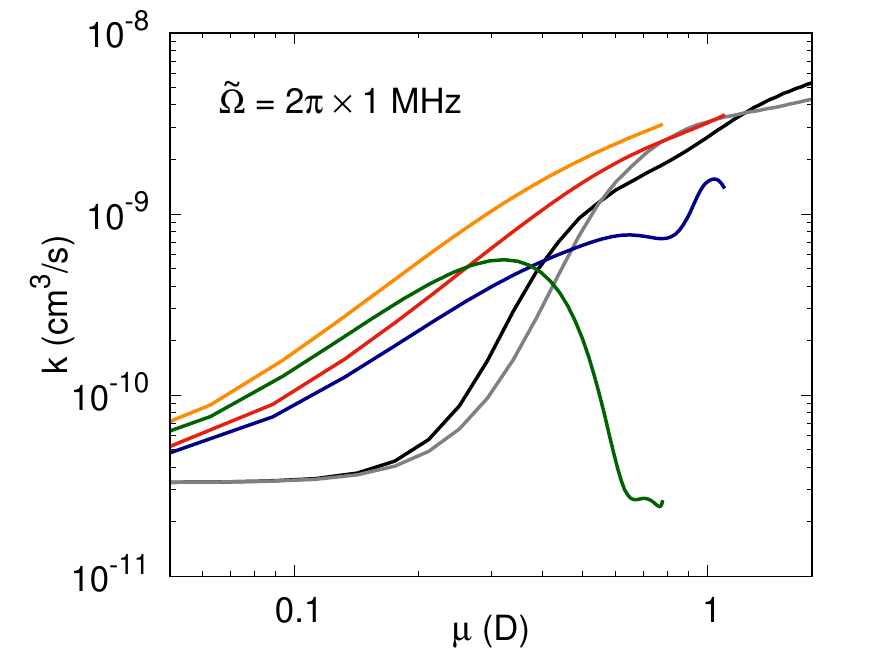}
\caption{ \label{fig:log}
Loss rates for fermionic NaK molecules polarized by microwave fields with $\tOmega=2\pi \times 1$~MHz, on a doubly logarithmic scale.
The data shown is identical to that in Fig.~\ref{fig:fermion_compare}(b),
but the logarithmic scale emphasizes off-resonant dressing which results in small induced dipole moments in individual molecules, yet causes resonant loss an order of magnitude faster than in the static case.
Red and orange lines show results at red detuning for linear and circular polarization, respectively.
Blue and green lines show results at blue detuning for linear and circular polarization, respectively.
For comparison, this graph includes the static field loss rates for the physical (sign-reversed) dipole-dipole interaction in black (gray).
}
\end{center}
\end{figure}

To further emphasize the qualitative differences between the first-order dipolar interaction -- accessible using static electric fields -- and the resonant dipolar interactions induced by microwave dressing at all realistically achievable Rabi frequencies,
Fig.~\ref{fig:log} shows again observable loss rates as a function of the induced dipole moment for $\tOmega=2\pi \times 1$~MHz, but now on a doubly logarithmic scale.
This emphasizes the regime of off-resonant dressing where the induced dipole moment is small.
Here, the first-order dipolar interactions are weak and their effect on the observable loss rates is small,
such that these are close to their universal value.
By contrast, the resonant dipolar interactions realized by microwave dressing lead to an order of magnitude higher collision rates.

\begin{figure*}
\begin{center}
\includegraphics[width=0.475\textwidth]{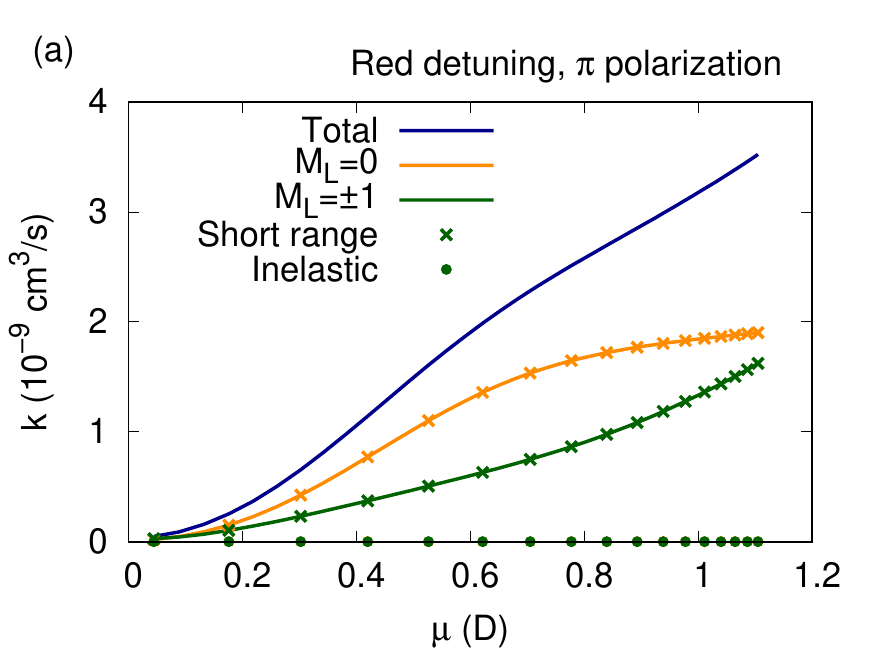}
\includegraphics[width=0.475\textwidth]{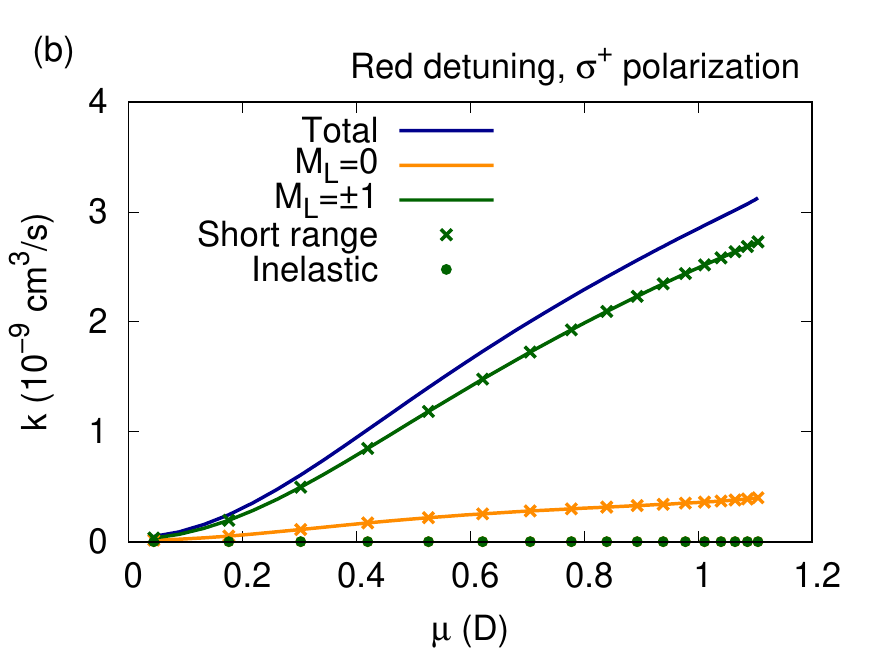}
\includegraphics[width=0.475\textwidth]{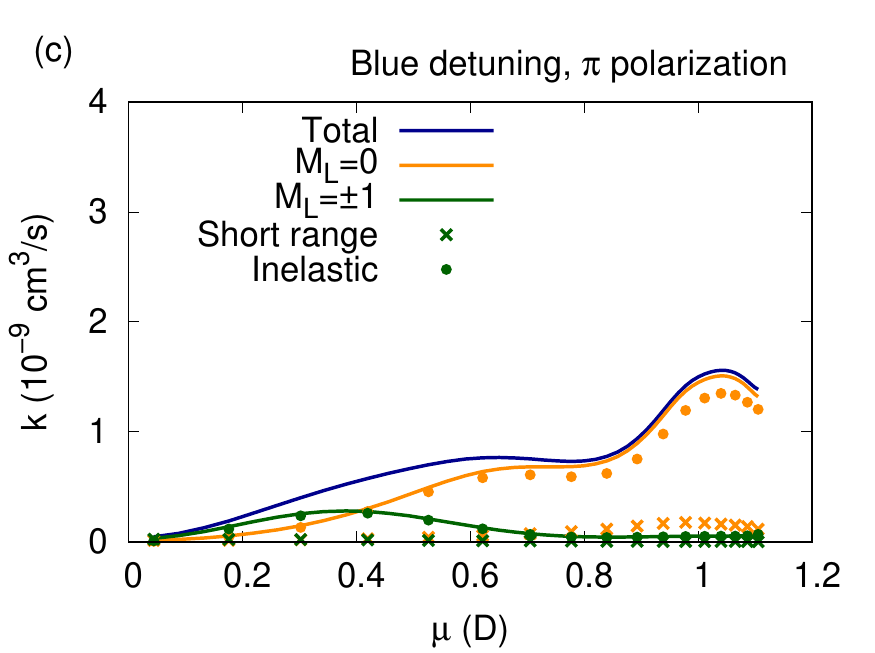}
\includegraphics[width=0.475\textwidth]{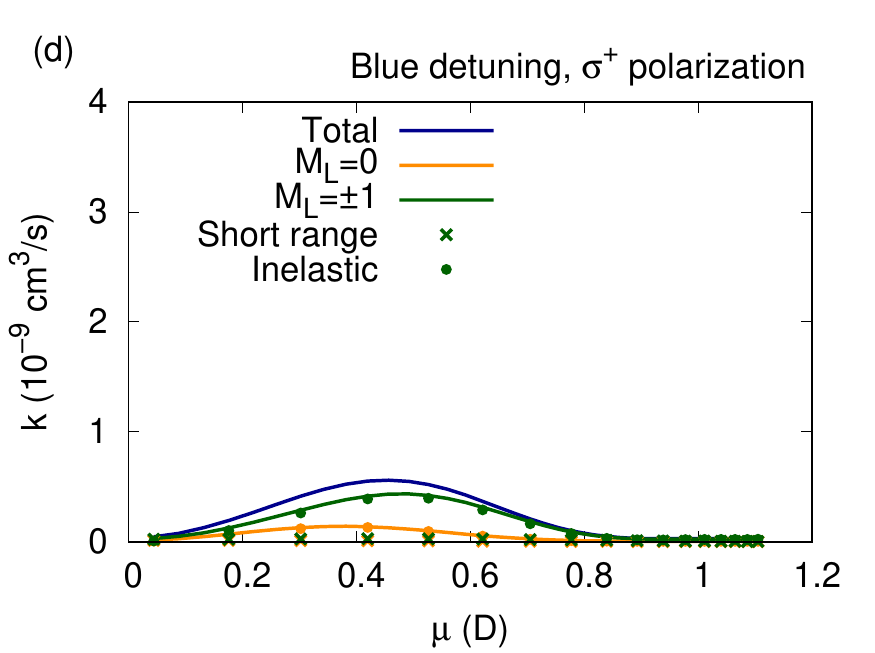}
\caption{ \label{fig:fermion_MW}
Losses for fermionic NaK molecules induced by microwave dressing with $\tOmega=2\pi\times 1$~MHz and (a) red detuning and $\pi$ polarization, (b) red detuning and $\sigma^+$ polarization, (c) blue detuning and $\pi$ polarization, and (d) blue detuning and $\sigma^+$ polarization.
Losses are decomposed into contributions of $M_L=0$ and $\pm 1$ (lines),
and these into short-range and inelastic losses, shown as correspondingly colored markers.
}
\end{center}
\end{figure*}

Figure~\ref{fig:fermion_MW} shows the $M_L$ decomposition of the loss curves in Fig.~\ref{fig:fermion_compare}(b), at $\tOmega=2\pi \times 1$~MHz.
For each $M_L$, the losses are further divided into those arising from reaching short range (crosses) and those due to inelastic scattering (dots).
For red detuning, shown in the top panels (a) and (b),
coupling to higher field dressed states leads to losses for all $M_L$ components,
even though dipolar interactions are repulsive for $M_L=\pm 1$ in the case of $\pi$ polarization [panel(a)] and $M_L=0$ for $\sigma^+$ polarization [panel(b)].
All losses are due to reaching short range, as inelastic processes are energetically not accessible.
If short-range loss can be eliminated, \emph{e.g.}, by turning off photo-excitation of collision complexes~\cite{christianen:19a,bause:21,gersema:21}, microwave dressing becomes a promising method for inducing strong interactions and large elastic cross sections, as argued in Ref.~\cite{yan:20}.
For blue detuning, shown in the bottom panels (c) and (d),
the losses are smaller as coupling to energetically lower field-dressed states induces repulsive interactions.
In this case, the losses are predominantly due to inelastic transitions to these lower field-dressed states.

\begin{figure}
\begin{center}
\includegraphics[width=0.475\textwidth]{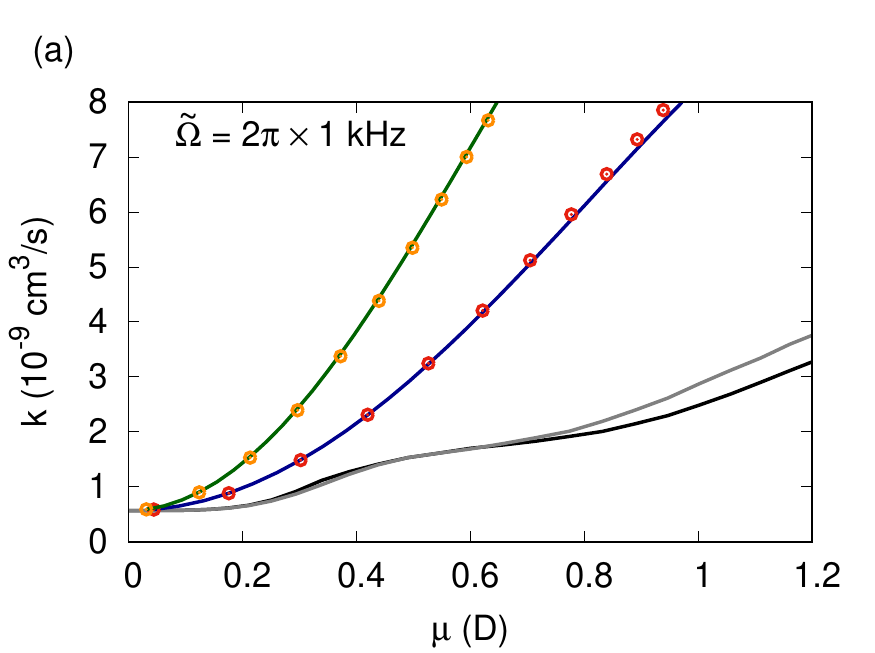}
\includegraphics[width=0.475\textwidth]{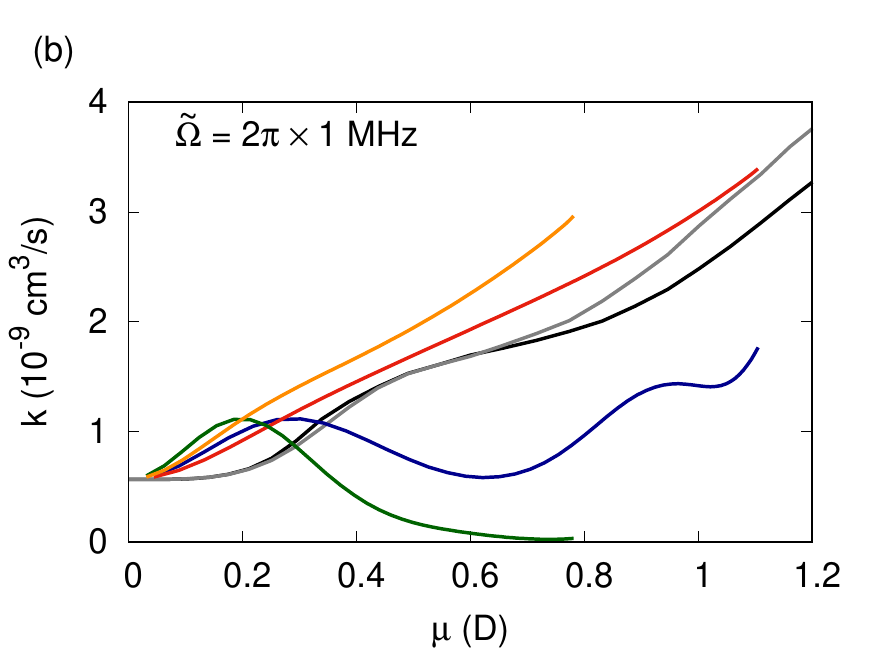}
\includegraphics[width=0.475\textwidth]{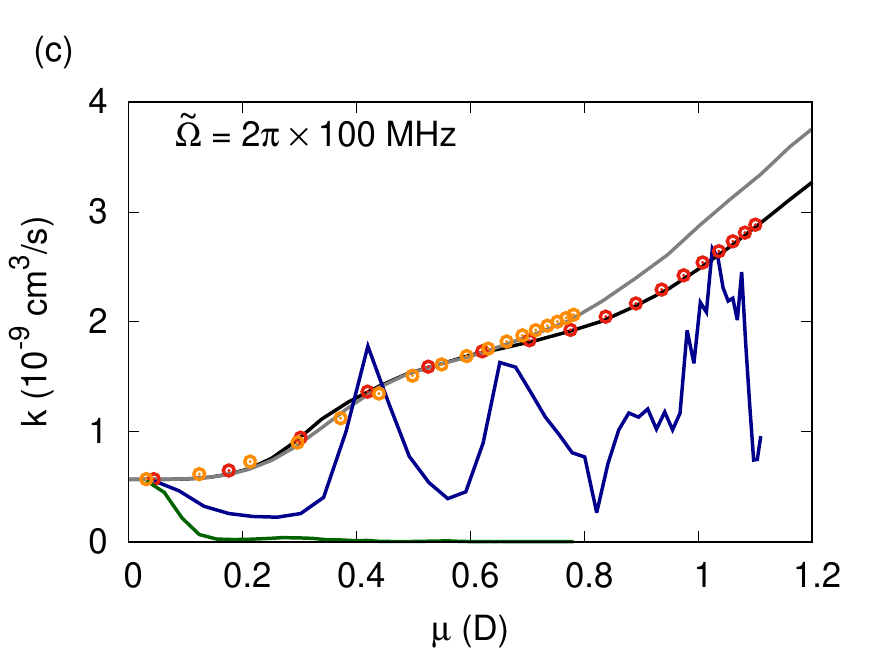}
\caption{ \label{fig:boson_compare}
Loss rates for bosonic NaK molecules polarized by microwave fields with (a) $\tOmega=2\pi \times 1$~kHz, (b) $\tOmega=2\pi \times 1$~MHz, (c) $\tOmega=2\pi \times 100$~MHz.
Red and orange lines shown results at red detuning for linear and circular polarization, respectively.
Blue and green lines show results at blue detuning for linear and circular polarization, respectively.
For comparison, this includes the static field loss rates for the physical (sign-reversed) dipole-dipole interaction in black (gray).
For clarity, some of the nearly overlapping lines are instead rendered as markers adhering to the same color coding.
	}
\end{center}
\end{figure}

\begin{figure*}
\begin{center}
\includegraphics[width=0.475\textwidth]{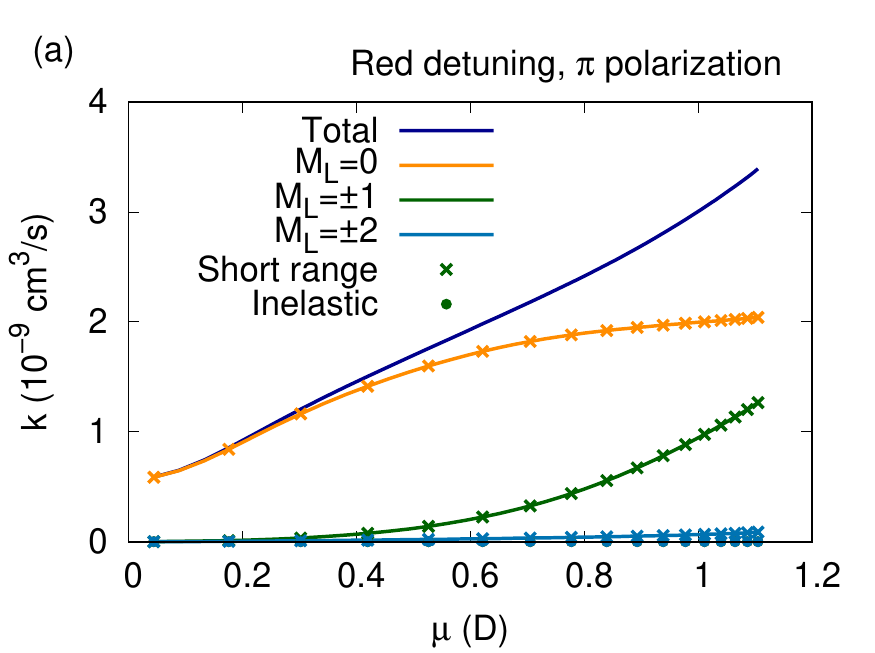}
\includegraphics[width=0.475\textwidth]{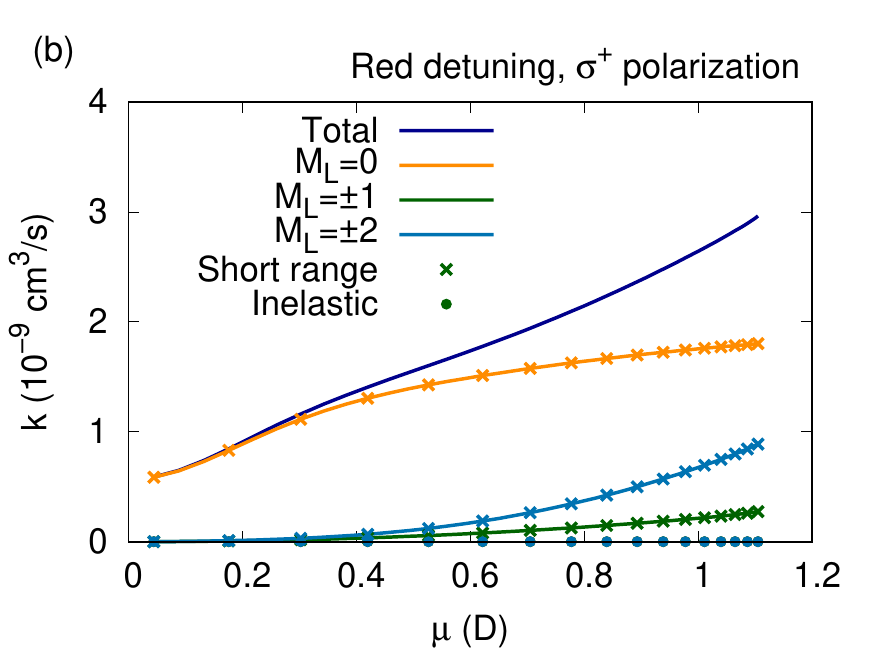}
\includegraphics[width=0.475\textwidth]{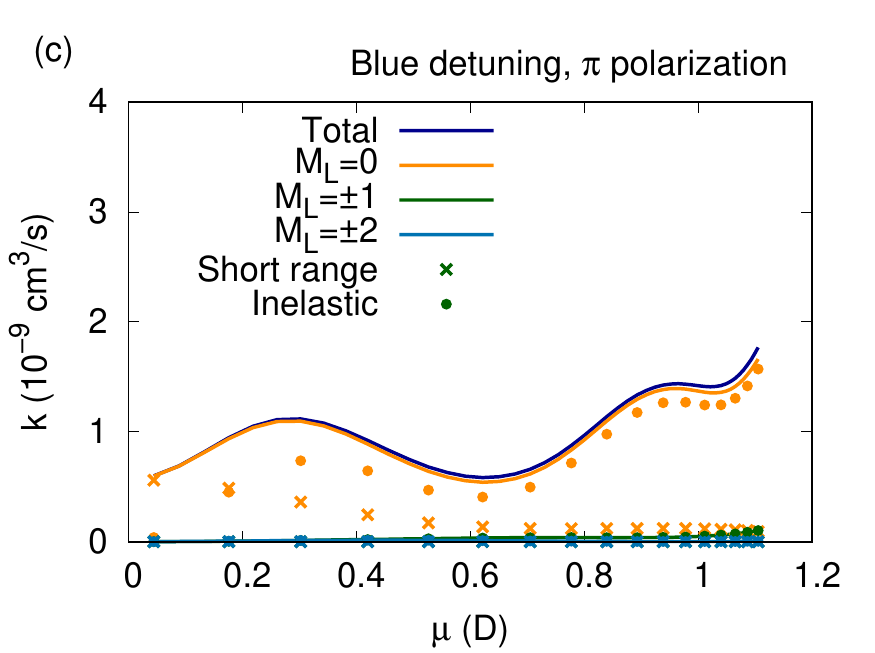}
\includegraphics[width=0.475\textwidth]{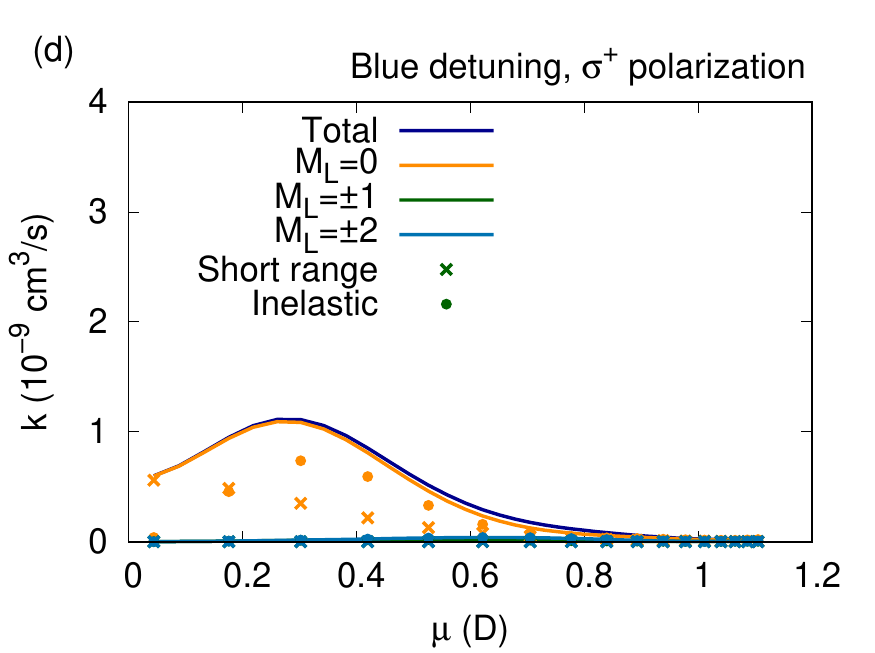}
\caption{ \label{fig:boson_MW}
Losses for bosonic NaK molecules induced by microwave dressing with $\tOmega=1$~MHz and (a) red detuning and $\pi$ polarization, (b) red detuning and $\sigma^+$ polarization, (c) blue detuning and $\pi$ polarization, and (d) blue detuning and $\sigma^+$ polarization.
Losses are decomposed into contributions of $M_L=0$, $\pm 1$, and $\pm 2$ (lines),
and these into short-range and inelastic losses, shown as correspondingly colored markers.
}
\end{center}
\end{figure*}

Figure~\ref{fig:boson_compare} compares loss rates for bosonic NaK molecules as a function of equivalent dipole moment for dipoles induced by static electric fields, and microwave electric fields with red or blue detuning and linear or circular polarization.
Panels (a), (b), and (c) show results obtained using $\tOmega=2\pi \times 1$~kHz, $\tOmega=2\pi \times 1$~MHz, and $\tOmega=2\pi \times 100$~MHz, respectively.
Similar to the fermionic case, we observe that losses of molecules dressed with red-detuned microwaves approach the space-fixed dipole approximation for high enough $\tOmega$.
For blue detuning, shielding can be obtained for circular polarization,
whereas losses structured with resonances are obtained for linear polarization.
Qualitative differences between the resonant dipolar interactions induced by microwave dressing and the space-fixed dipole approximation are observed for all polarizations and detunings for realistic Rabi frequencies.
Figure~\ref{fig:boson_MW} shows the decomposition into short-range and inelastic loss processes for different $M_L$.

\section{Conclusions \label{sec:conclusions}}

In conclusion, we have compared the dynamics of collisions between ultracold molecules with dipole moments induced in different ways:
by applying either a static electric field, or microwave fields with linear or circular polarization and blue or red detuned from the first rotational transition.
We find that the dynamics of ground state molecules polarized by a static field are accurately described by the ``space-fixed dipole'' approximation.
For molecules dressed with microwaves, the space-fixed dipole approximation is accurate only for red detuning and high Rabi frequencies in the order of 100~MHz.
For lower Rabi frequency or blue detuning,
the collision dynamics are dominated by resonant dipolar interactions that quantize the dipoles of colliding molecules along the intermolecular axis,
a qualitatively different scenario from first-order dipolar interactions between molecular dipoles oriented along an external field.
For blue-detuned microwaves,
the resonant dipolar interactions are repulsive,
which leads to substantial suppression of loss by microwave shielding for circular polarization~\cite{karman:18d,karman:19c,karman:20,anderegg:21},
or to loss rates that are structured by dipole-tunable resonances for linear polarization.
At more realistic Rabi frequencies on the order of 1~MHz,
resonant dipolar interactions are observed for all microwave polarizations and detunings.

\newpage

\end{document}